# Copper doping effects on the superconducting properties of Sm-based oxypnictides

Mohammad Azam[1], Manasa Manasa[1], Tatiana Zajarniuk[2], Taras Palasyuk[1], Ryszard Diduszko[3], Tomasz Cetner[1], Andrzej Morawski[1], Cezariusz Jastrzebski[4], Michał Wierzbicki[4], Andrzej Wiśniewski[2], Shiv J. Singh[1*]

[1]*Institute of High Pressure Physics (IHPP), Polish Academy of Sciences, Sokołowska 29/37, 01-142 Warsaw, Poland*

[2]*Institute of Physics, Polish Academy of Sciences, aleja Lotników 32/46, 02-668 Warsaw, Poland*

[3]*Łukasiewicz Research Network Institute of Microelectronics and Photonics, Aleja Lotników 32/46, 02-668 Warsaw, Poland*

[4]*Faculty of Physics, Warsaw University of Technology, Koszykowa 75, 00-662 Warsaw, Poland*

*\*Correspondence address:* [sjs@unipress.waw.pl](mailto:sjs@unipress.waw.pl)

[https://orcid.org/0000-0001-5769-1787](https://orcid.org/0000-0001-5769-1787)

**Abstract**

A systematic investigation has been performed by synthesis and comprehensive characterization of a series of $SmFe_{1-x}Cu_xAsO_{0.8}F_{0.2}$ bulks ($x = 0$ to $0.2$). These samples are well characterized by structural, Raman spectroscopy, microstructural, transport, magnetic measurements, and supplementary calculations within density functional theory (DFT). The parent compound, $SmFeAsO_{0.8}F_{0.2}$ (Sm1111), exhibits a superconducting transition temperature ($T_c$) of approximately 54 K. The lattice volume ($V$) is increased with Cu substitution ($x$) without observing any impurity phase related to copper, which confirms the successful incorporation of Cu at Fe sites in the superconducting FeAs layers. These analyses are also well in agreement with Raman spectroscopy measurements and relevant DFT results. The superconducting transition is decreased systematically with copper doping and completely suppressed for 7% Cu-doped Sm1111 ($x = 0.07$). A large amount of Cu substitution ($x \geq 0.07$) has demonstrated the metal to insulate transition in the low-temperature range, and no impurity phase was observed even at high Cu doping levels ($x = 0.2$). The calculated critical current density of the parent sample is suppressed with copper substitution, suggesting the reduced pinning centers, sample density, and grain connections, as confirmed by the microstructural analysis. Our studies suggest that the substitution of Cu in the superconducting FeAs layer, resulting the enlargement of the lattice volume, is a source of strong disorder scattering, leading to the suppression of $T_c$ and the emergence of metal-to-insulator, unlike the more successful carrier doping by nickel (Ni) or cobalt (Co), as previously reported.

## Introduction:

In 2008, iron-based superconductor (FBS) was discovered through F-doped LaFeAsO with a critical transition temperature ($T_c$) of 26 K [1]. Following this groundbreaking report, many compounds belonging to this high $T_c$ superconductor have been discovered [2, 3]. $RE$FeAsO (1111; $RE$ = rare earth) is a very promising and unique family of iron-based superconductors [4, 2] and provides the highest transition temperature of 58 K [5, 6]. The parent compound $RE$FeAsO does not depict the superconductivity and is a semimetal that undergoes a structural distortion from tetragonal (space group *P4/nmm*) to orthorhombic (space group *Cmma*) symmetry at ~155 K [1, 7], and further lowering the temperature, a paramagnetic to spin density wave (SDW) or a long-range antiferromagnetic transition appears around 135 K [8, 9]. The oxypnictide 1111 has a layered tetragonal ZrCuSiAs type structure with a space group of *P4/nmm* [10, 11], consisting of two-dimensional layers of ($RE$-O) and (Fe-As). In this structure, the rare earth $RE$ ions are coordinated by four arsenic and four oxygen ions, forming a distorted square antiprism, while iron ions are tetrahedrally coordinated by four As ions [1]. The superconducting Fe-As and $RE$-O layers are the conducting layers and charge reservoirs, respectively [12, 11]. The substitution of fluorine at oxygen sites or oxygen deficiency suppresses this anomaly, and the superconductivity appears [13]. Replacement of La with the smaller rare earth metals such as Ce, Pr, Nd, Sm, Gd, Tb, and Dy [14] enhances the superconducting transition temperature and reaches a maximum $T_c$ of 58 K in SmFeAsO$_{1-x}$F$_x$ ($x = 0.25$) [5]. Very few studies have been reported to induce the superconductivity by the hole doping in these oxypnictides, such as Sr-doped LaFeAsO at 25 K [15].

Iron-based superconductors have many electron and hole pockets in the Brillouin zone and are recognized as multiband superconductors [16, 17, 18], where electron and hole doping are common methods to induce the superconducting properties. Generally, Fe-As-Fe bond angles and the Fe-As tetrahedron geometry play a crucial role in determining the electronic and magnetic properties of this 1111 oxypnictide, and a high superconducting transition is observed when the lattice distortion reaches the minimum level [19]. The dopants having a low ionic radius reduce the lattice distortion and can be quite effective in enhancing the superconducting transition temperature, as reported for various FBS [2]. Hence, the substitution plays an important role in tuning the superconducting properties of FBS, and many kinds of doping have been reported, such as F doping at O-sites [2], Co, Mn, and Ni doping at Fe-sites [20, 21, 22], and Sb doping at As-sites [23], to understand the physical and magnetic properties of iron-based superconductors [24]. Interestingly, FBS shows some completely different behaviors compared to cuprate superconductors [3], such as the substitution of a transition metal such as Co at Fe-sites, which also induces the superconductivity, whereas the substitution of any transition metal ion for copper drastically lowers the $T_c$ [20] [25].

A lot of unique characteristics are observed through the transition metal substitution at the Fe site in FBS. Superconductivity, for instance, can be induced by Co and Ni doping in the 1111 family [21, 20, 25]. Doping by many transition metals such as Co, Mn, Ru, and Ni has been reported for the 1111 family, and a systematic evaluation of superconducting properties has been observed [26, 21, 24]. A study based on isovalent Ru-doped Sm1111 shows a notable reduction of $T_c$ and suggests that intraband scattering is enhanced due to the disorder induced by the dopants, whereas no significant effect is observed for the interband scattering [26]. This reduction of superconducting properties by Ru-doping is due to either disorder in the Fe sub-lattice or the appearance of a short-range magnetic order [26]. Similar to this, the reduction of superconducting properties is also reported for Mn and Ni doping in Sm1111 [21]. In all these transition metal doping, Cu doping plays an interesting role in the superconducting properties of FBS, as reported for $BaFe_2As_2$ (122) [27], FeSe (11) [28], and LiFeAs (111) materials [29]. Cu substitution in FeSe bulks, i.e., $Fe_{1-x}Cu_xSe$, destroys the superconductivity at a very low level ($x \sim 1.5\%$), and a metal-insulator transition appears at 4% Cu dopants ($x \sim 0.04$), which is different from Co and Ni-doped FeSe and Cu-doped 122 [28, 22, 30]. Theoretical calculations suggest that Cu doping acts as a source of strong disorder scattering due to a nominal $d^{10}$ configuration instead of a $d^9$ one, which results in a reduction of transition temperature and the appearance of a metal-to-insulator transition. On the other hand, the density functional studies suggest that Cu doping serves as an electron dopant source, but it still plays a role in strong scattering [30]. Studies based on $Ba(Fe_{1-x}Cu_x)_2As_2$ and $NaFe_{1-x}Cu_xAs$ from angle-resolved photoemission spectroscopy (ARPES) reveal the localization of the part of the electron due to the Cu substitution [27, 31, 32]. Currently, the investigation of Cu doping is mostly focused on the 122, 11, and 111 families. Until now, there has been no report of Cu doping in the 1111 family, which is the main motivation behind this research work.

This study focuses on systematically investigating the superconducting characteristics of a series of Cu-doped SmFeAs(O,F) polycrystalline samples that are produced using a one-step solid-state reaction approach. We find that the lattice volume ($V$) is increased and the sample density is slightly reduced with Cu doping. Transport and magnetic measurements depict that the transition temperature and critical current are decreased with Cu-substitutions, as well as accompanied by a broadening of the transition width. This effect is likely caused by strong disorder scattering rather than a change in carrier density resulting from the addition of Cu to the superconducting layer. Additionally, the metal-to-insulator transition is observed at large levels of Cu-doping, which is likely the result of the previously reported Anderson localization of charge carriers as a result of disorder induced by Cu-doping [33].

## Experimental:

The initial precursors, Sm powders (99.9%), As chunks (99.99%), Cu powder (99.9%), Fe powder (99.99%), $Fe_2O_3$ powder (99.85%), and $FeF_2$ powder (99%), were used to synthesize bulk $SmFe_{1-x}Cu_xAsO_{0.8}F_{0.2}$ by following the one-step solid-state reaction method [5]. SmAs was initially prepared by reacting Sm and As powder at 500°C for 15 hours. These materials were mixed according to the stoichiometric formula $SmFe_{1-x}Cu_xAsO_{0.8}F_{0.2}$ ($x$ = 0, 0.01, 0.02, 0.03, 0.04, 0.05, 0.07, 0.10, and 0.20), grounded in a mortar pestle, and compacted into disk pellets (diameter: 10 mm) by the hydraulic pressing at ~200 bars. Those pellets were placed in a tantalum (Ta) tube as a crucible and sealed in an evacuated quartz tube, which was heated at 900°C for 45 hours [5, 23]. All growth processes were performed in an inert gas glove box with very low oxygen and moisture levels (< 1 ppm).

The room-temperature powder X-ray diffraction (XRD) measurements were used for the structural analysis of these samples and performed by a Rigaku SmartLab 3 kW diffractometer with Bragg Brentano configuration, filtered Cu-Kα radiation (wavelength: 1.5418 Å, power: 30 mA, 40 kV) and a Dtex250 linear detector with the measured profile from 5° to 70° with a step rate of 0.01°/min. Rigaku's PDXL software and the ICDD PDF4+2023 standard diffraction patterns database were employed for profile analysis and lattice parameter calculations. Microstructural analysis and elemental mapping were conducted using the Zeiss Ultra Plus field-emission scanning electron microscope equipped with energy-dispersive X-ray analysis (EDAX). Magnetic measurements were carried out using vibrating-sample magnetometry (VSM) attached to Quantum Design PPMS under the magnetic fields up to 9 T in the temperature range of 4-60 K. The magnetic susceptibility was measured in zero-field-cooled (ZFC) and field-cooling (FC) modes in the presence of a magnetic field of 20 Oe. The temperature dependence of resistivity measurements was conducted using a four-probe method attached to a closed-cycle refrigerator (CCR) over a temperature range of 7-300 K at zero magnetic field. Furthermore, the sample homogeneity and their impact on the superconducting properties of Cu-doped Sm1111 bulks have been verified by means of a number of experiments conducted on the two pieces (Sample-1 and Sample-2) of each Cu-doped sample from the same batch. In relation to it, the measurements and analysis of a few chosen samples ($x$ = 0, 0.01, 0.02, 0.03, 0.04, and 0.20) are shown in Figures S2, S3, S4, S6, S7, S8, and S9 in the supplemental file.

Raman scattering measurements were performed using LabRam ARAMIS (Horiba Jobin Yvon) spectrometer. Samples were excited by visible light with a 632.8 nm wavelength of an ion He-Ne laser. Incident light was focused by a 100x objective with NA 0.95 to a spot of < 3 μm on a probed sample. The same objective was used for collecting backscattered light that was further dispersed by a diffraction grating of 2400 l/mm and further registered by a charge-coupled device (CCD), yielding Stokes part of Raman shifts in the range from 65 to 300 cm$^{-1}$. Laser power was attenuated to a few mWatts in order to avoid sample overheating. Raman spectra were acquired at room temperature with an acquisition time of 120 s or longer, averaged by 3 accumulations. Measurements were done on the sample surface as provided after synthesis without any additional treatment. As a consequence of the polycrystalline nature of the samples, the collected spectra showed a sizable scatter of the measured data as regards the intensity and position of detected peaks, depending on a point across the sample being examined. These observations might be ascribed to different grain orientations and lattice strains induced by Cu substitution. Therefore, we did not perform an analysis of doping-induced effects on peak intensity. For each content of Cu substitution, Raman spectra were collected on average from at least ten points on the sample to ensure sufficient statistics as far as peak positions were concerned. Signals in the measured spectra were deconvoluted by peaks of Lorentzian shape.

## Results and discussion:

The purity and crystallinity of the prepared $SmFe_{1-x}Cu_xAsO_{0.8}F_{0.2}$ polycrystalline samples are systematically investigated through powder X-ray diffraction (XRD) measurements. In Figure 1(a), the room-temperature XRD patterns of $SmFe_{1-x}Cu_xAsO_{0.8}F_{0.2}$ polycrystalline samples are presented, corresponding to different Cu doping levels ($x$ = 0, 0.01, 0.02, 0.03, 0.04, 0.05, 0.07, 0.10, and 0.20). The crystal structure observed in the XRD patterns is consistent with the tetragonal ZrCuSiAs-type structure with the space group *P4/nmm*, as similar to the previous reports [5, 10]. The parent compound, $SmFeAsO_{0.8}F_{0.2}$ bulks, reveals the presence of impurity phases such as the SmOF phase and a tiny amount of SmAs, and has the obtained lattice parameters $a$ = 3.929(9) Å and $c$ = 8.502(6) Å for $x$ = 0. This observation aligns with prior reports and is attributed to the synthesis process [5, 23]. However, it is noteworthy that with increasing Cu doping levels, both the quantity and percentage of impurities are almost the same as those observed for the parent $SmFeAsO_{0.8}F_{0.2}$ sample [5]. All samples also have a tetragonal structure, as shown in Figure 1(a). Figure 1(b) depicts a clear shift of the main peak (102) of the Sm1111 phase with respect to copper substitution contents and suggests successful Cu doping induced inside the superconducting Sm1111 lattice. The calculated lattice parameters ($a$, $c$) and unit cell volume ($V$) for these samples are shown in Figure 1(c)-(e). We have also included the data based on Ni and Ru-doped SmFeAs(O,F) from the previous reports [21, 26]. Since Cu has a bigger ionic radius than that of iron, the lattice parameter '$a$' (Figure 1(c)) increased with the nominal value of Cu-doping ($x$). This behavior is almost similar to that of Ni and Ru-doped Sm1111 [21, 26]. However, an almost constant lattice parameter '$c$' is observed for Cu doping, which is nearly the same behavior for the isoelectronic Ru substitution at the Fe site and for Cu-doped LiFeAs [29]. Whereas Ni doping has a much larger effect by shrinking the parameter '$c$', as shown in Figure 1(d). Interestingly, the overall tetragonal lattice expands due to the Cu, Ru, and Ni doping, which is clear from the variation of cell volume '$V$' with the nominal doping level, as shown in Figure 1(e). Ni substitution induces the additional electron, as shown by ARPES [34], whereas Ru acts as an isoelectronic substitute. On the other hand, the 3d electrons from the Cu dopant are highly confined and have minimal contribution to the Fermi surface, as observed in the case of Cu doping in LiFeAs and BaFe$_2$As$_2$ [27, 29]. The replacement of Co or Ni adheres to a rigid band model due to the overlapping of the d-bands of Co and Ni with the Fe d-band, resulting in a lack of distinguishing characteristics compared to the Fe d-band. On the other hand, Cu has a deeper impurity potential [29], which causes most of the Cu 3d orbital electrons to become localized. When compared to Co/Ni substitution, this is compatible with a much smaller contraction of the lattice parameter '$c$' by Cu substitution, which leads to the generation of itinerant electrons [29]. Furthermore, a study of Cu-doped LiFeAs also suggests a small electron doping effects due to Cu substitution and the effect of disorder with strong impurity potential [29]. These reported investigations indicate that Cu substitution results in substantial impurity scattering rather than a change in carrier density. Interestingly, Cu doping acts in a similar way as that of Ru doping, but the mechanism is different [34]. One can note that charge carriers are localized in the case of Cu-doping, but no extra carriers are induced in the case of Ru-doping. As a result, the effects of Cu and Ru-doping are nearly identical, as evidenced by the lattice parameters and cell volume [26], as shown in Figure 1(c)-(e).

The room-temperature unpolarized Raman spectra of Cu-doped SmFeAsO$_{0.8}$F$_{0.2}$ in backscattering configurations are compared in Figure 2 with respect to various Cu substitutions ($x$). Raman spectra of pristine SmFeAsO$_{0.8}$F$_{0.2}$ revealed the presence of signals related to Sm(A$_{1g}$), As(A$_{1g}$), and Fe(B$_{1g}$) out-of-plane lattice modes, as shown in Figure 2(a), consistent with earlier observations reported in the literature [35, 36, 37]. For copper-doped Sm1111 samples, *i.e.*, $SmFe_{1-x}Cu_xAsO_{0.8}F_{0.2}$ ($x = 0$, 0.07, and 0.20), no major changes in the Raman spectra are detected that indicate the stability of crystal structure as well as rather unchanged chemical bonding upon copper substitution. As regards peak positions, a considerable softening (shift to lower energy) of the Fe-related mode and a similar but less pronounced change in the energy of the As-based mode were detected upon increasing copper content, constituting a decrease in Raman frequency of ca. 6 $cm^{-1}$ and 2 $cm^{-1}$, respectively. Within the experimental error, no considerable change in the energy of the Sm-related mode was observed.

The phonon properties of $SmFe_{1-x}Cu_xAsO_{0.8}F_{0.2}$ for copper doping ($x = 0$ (parent), 0.5, and 1.0 (full substitution)) were assessed by density functional theory (DFT) calculations as supplementary to the experimental Raman scattering evaluation presented in this study. Structure geometries and atomic forces were determined using the all-electron full-potential linearized augmented plane-wave (LAPW) program WIEN2k [38]. Perdew, Burke, and Ernzerhof (PBE) form of the generalized gradient approximation (GGA) to the exchange-correlation function [39] was utilized. Plane-wave cut-off parameter $Rk_{max}$ was set to 8, along with 1000 k-points in the first Brillouin zone and 6 nonspherical matrix elements for large spheres (LVNS). The criteria for self-consistent field (SCF) convergences were considered as: charge 0.0001 *e* and energy 0.0001 Ry. The rest of the initialization parameters were set to default values, and the lattice constants for SmFeAsO were taken from Ref [36]. For the SmCuAsO compound, the lattice constants were determined by energy optimization of cell volume, with subsequent optimization of the lattice parameter ratio (*c/a*). For mixed compounds with fractional concentrations (*x*) of Cu atoms, the lattice constants of supercells were determined by linear interpolation between $x = 0$ (pure Fe) and $x = 1$ (pure Cu). The experimental lattice parameters determined by the XRD analysis have almost similar behavior to the calculated lattice parameters from DFT calculations. The variation of the calculated and experimental values of the lattice parameters ('*a*' and '*c*') is depicted in the supplementary Figure S1(a)-(b). Although the lattice constant '*c*' is underestimated by 0.3 Å, the ab initio findings provide an appropriate trend with regard to copper concentration for both lattice parameters. One can note that DFT calculations are known to underestimate or overestimate basic parameters such as lattice constants and band-gaps [40]. In each case of Cu-substituted Sm1111, the internal free parameters of Wyckoff positions were optimized by minimization of internal forces with the SCF convergence criterion of 0.1 mRy/a.u. The frequencies of Raman modes were determined by the frozen-phonon method. Symmetric Raman-active modes were determined by a group theoretical method [41]. Force constants were obtained by the first derivative of the 5-point cubic interpolation of the force vs. atom displacement relationship. Our DFT calculations also revealed considerable softening of all Raman modes for $SmFeAsO_{0.8}F_{0.2}$ with increasing doping levels, as depicted in Figure 2(b) and observed in the experiment. Similar to the experimental observations, the Fe-related mode was found to show the highest rate of softening if compared to vibrations of As. The frequency of Sm-related mode was found to be the least sensitive to doping. According to calculations, if doping increased from $x = 0$ to $x = 1$ (fully substituted compound), the force constants of Fe (Cu), As, and Sm vibrations decreased by 62.8%, 26.1%, and 15.6%, respectively, which may plausibly account for the different rates of softening yielded by theory and observed in our experiment. We have also analyzed the possible lattice distortion induced by Cu doping based on DFT data, as shown in Figure 2(c). In the parent compound ($x = 0$), the Fe-As distances within the FeAs layer are uniformly approximately 2.32434 Å. Nevertheless, the incorporation of 12.5% ($x = 0.125$) and 25% ($x = 0.25$) Cu-doping induces a significant deformation in the crystal structure of the FeAs layer, leading to the existence of Fe-As chemical bonds with different bond lengths. These enhanced Fe-As distances will also affect the corresponding Fe-As-Fe bond angles, as reported for FBS [3, 19]. FBS system that has an ideal Fe-As-Fe bond angle of 109°47' for the perfect FeAs tetragonal structure can depict the highest transition temperature. Hence, the deviation of this Fe-As-Fe bond angle from this ideal value can distort the FeAs tetragonal lattice, and a reduction of the superconducting transition can be observed [19]. Based on these reports [3, 19], our DFT analysis suggests that Cu doping produces a distortion inside the Sm1111 crystal structure.

The elemental mapping for various bulk samples is performed and illustrated in Figure 3(a), (b), (c), and (d) for the selected samples $x = 0$, 0.02, 0.04, and 0.2, respectively, whereas the mapping of other samples is depicted in Figure S3, S4, S5. Two pieces of samples (Sample-1 and Sample-2) from the same batch for the sample $x = 0$, 0.02, 0.04 and 0.2 are depicted as Figure 3 and Figure S2 with respect to Sample-1 and Sample-2, respectively. The parent sample has almost a homogeneous distribution of all constituent elements (Sm, Fe, As, O, and F), and very few places are observed with rich Sm, O and F, suggesting the possibility of the SmOF ($Sm_2O_3$) phase, as also observed from the XRD measurement. A small amount of copper doping, such as $x = 0.01$, has areas of rich Sm, As, and O/F elements compared to the parent sample, suggesting the existence of SmAs and SmOF phases, but the distribution of Cu elements is almost homogeneous in the sample, as illustrated in Figure S3(a)-(b). Furthermore, the sample $x = 0.02$ has many small areas that are inhomogeneous with the constituent elements Sm, O, and F (Figure 3(b) and S2(b)), which confirms the existence of the SmOF and SmAs phases, as also confirmed from Figure 1. Further increase of Cu substitutions, i.e., $x = 0.03$ sample seems porous, and almost the same inhomogeneous distribution of Sm, O, and F elements is observed (Figure S4(a)-(b)) as that of $x = 0.02$. Few areas are detected with rich Sm, O, F, Fe, and As elements (Figure 3(c) and S2(c)) for sample $x = 0.04$, suggesting the presence of SmOF and FeAs phases, and following the XRD analysis. Further increases of Cu doping contents, *i.e.,* for $x = 0.05$, lead to a large pore size where almost no elements were observed (Figure S5(a)), and also the SmOF phase does exist. The sample $x = 0.1$ (Figure S5(b)) has many pores and an almost homogeneous distribution of the constituent elements, but in a few places SmOF and SmAs/FeAs phases are detected. The sample with a large amount of Cu doping, i.e., $x = 0.2$ (Figure 3(d) and S2(d)), looks compact and has an almost homogenous distribution of the elements with the possibility of SmOF phase, which seems comparable to the parent compound (Figure 3(a)). These mapping analyses suggest a homogeneous distribution of Cu dopants and confirm the presence of SmOF and SmAs phases for all Cu-doped Sm1111, which aligns well with the previously discussed XRD data.

Figure 4 depicts the backscattered electron (BSE) images for the parent ($x = 0$) and Cu-doped Sm1111 ($x = 0.02, 0.04, 0.20$) bulks. These polycrystalline samples were polished with various grades of emery paper without using any liquid or oil. White, gray, and black contrasts are observed for $Sm_2O_3$ (SmOF), $SmFe_{1-x}Cu_xAsO_{0.8}F_{0.2}$, and pores (SmAs or FeAs), respectively. BSE images of two pieces of samples (Sample-1 and Sample-2) from the same batch for the sample $x = 0, 0.02, 0.04,$ and $0.2$ are depicted as Figure 4 and Figure S6 corresponding to Sample-1 and Sample-2. The parent sample $x = 0$ has a compact microstructure, but the white contrast and many small pores are observed, as depicted in Figure 4(a)-(b). Furthermore, many grains do not seem to be well-connected, and the grain size is observed to be around 1-2 µm (Figure 4(c)). Interestingly, a small amount of Cu substitution in Sm1111 seems to reduce the grain connectivity and sample density, as clear from Figure 4(d)-(f), and contains many micro- or nanopores. A further increase of Cu-substitution, such as $x = 0.04$, seems to slightly improve the sample density and grain connectivity by reducing the number of pores, as observed from Figure 4(g)-(i), but still there exist nanopores in between many grains. The phases $Sm_2O_3$ (SmOF) and FeAs are observed at a few places, as revealed in Figure 4(g), which is well in agreement with the elemental mapping and XRD data analysis. However, the samples $x = 0.05$ and $0.10$ again look very porous, as not depicted here, and many nanopores are observed and grain size is reduced. Furthermore, a large amount of Cu doping, i.e., the sample $x = 0.2$, has many nanopores, but similar to other samples, SmOF or $Sm_2O_3$ and also SmAs/FeAs phases are detected at a few places, as depicted in Figure 4(k)-(l). These nanopores exist between grains (Figure 4(l)) and reduce grain connections and grain size, similar to the other Cu substitution sample. These analyses suggest that a very low amount of Cu-doped samples (x<0.05) reduces the compactness of the samples compared to that of the parent compound, which could be due to the presence of many micro- and nanopores, and increasing the number of pores. Hence, it suggests that very low amounts of Cu-doped samples have lower connectivity between Sm1111 grains and a lower density than that of the parent sample. The number of nanopores is increased for very large amounts of Cu-doped samples, as evidenced from Figures 4(j) and 4(l). All Cu-doped samples have an almost homogeneous microstructure with the existence of a small amount of SmOF/$Sm_2O_3$ phase similar to the parent sample. Interestingly, the number of micro- and nanopores is increased for the Cu-doped samples, which exists between grains and reduces the grain connectivity and grain size, as clearly observed from the high-resolution images, and the samples seem porous with Cu-substitution compared to the parent $x = 0$. By considering the theoretical density of 7.1 g/cm$^3$ for Sm1111 [3], the density for the parent compound is around 51%, whereas Cu-doped Sm1111 has 48%, 49%, 46.3%, 48%, and 49% for $x = 0.01, 0.02, 0.05, 0.1,$ and 0.2, respectively. It suggests that the sample density is slightly reduced with Cu substitution compared to the parent compound. This is well in agreement with the above-discussed analysis of the microstructure and elemental mapping.

The temperature dependence of the resistivity ($\rho$) behavior for various Cu-doped SmFeAs(O,F) bulks is illustrated in Figure 5(a). The resistivity measurements of two pieces of sample (Sample-1 and Sample-2) $x = 0, 0.02, 0.04,$ and 0.2 from the same batch are shown in Figure S7 for Sample-1 and Sample-2, respectively. The parent compound exhibits a linear normal resistivity variation and has a low resistive value across the whole temperature range. The resistivity value is increased for $x = 0.01$ and $x = 0.02$ compared to that of the parent sample. It could be possible due to the reduced grain connections, i.e., the existence of many pores. According to the previous report [29], Cu substitution produces disorder or defects inside the lattice and contributes to the scattering centers of the carriers, which could be a reason for the high value of $\rho$ of Cu-doped Sm1111 compared to that of the parent compound. Further increase of Cu doping reduces the normal state resistivity as evidenced by the results obtained for the sample with $x$ values of 0.03, 0.04, and 0.05. The superconducting transition is systematically reduced with Cu doping. More than 5% of Cu-substitution in F-doped Sm1111 samples exhibit an absence of the superconducting transition within the measured temperature range of up to 7 K. A linear decrease of the normal resistivity is observed for the sample $x = 0.07$, but in the low-temperature range, a negative slope of the resistivity emerges, i.e., a metal-to-insulator transition appears, similar to the reported Cu-doped FeSe [28] and LiFeAs [29]. This effect is more pronounced for Sm1111 with increased Cu doping, i.e., specifically for $x = 0.1$ and 0.2, as depicted in the inset of Figure 5(a). The resistivity value of the low amount of Cu-doped samples in the normal state is not consistent, which could be possible due to the non-homogeneous distribution of the constituent elements into Cu doped Sm1111 caused by Cu dopants. When the disorder scattering due to Cu-doping in FBS is so strong, it significantly reduces the electron mobility and detriments the superconductivity properties [30]. Because of this possible reason, no superconductivity is detected here for more than 5% Cu substitution in $SmFeAsO_{0.8}F_{0.2}$, and a metal-to-insulator transition is observed that generally follows Anderson localization effects, resembling the phenomenon as reported for Cu-doped FeSe [30, 33].

The low-temperature resistivity behavior is depicted in Figure 5(b) for these Cu-doped Sm1111 samples. The parent compound possesses an onset transition ($T_c^{onset}$) of 53.8 K and an offset transition ($T_c^{offset}$) of 48.4 K, which is similar to the previous reports [5]. The method of determining the transition temperatures ($T_c^{onset}$ and $T_c^{offset}$) is mentioned in the inset of Figure 5(b) and is applied to all these samples. The transition temperature is significantly reduced just for 1% copper-doped Sm1111, and the onset $T_c$ is observed around 40 K with the increased transition width ($\Delta T$), which is consistent with the findings reported for Sm1111 with a small amount of Sb substitution [23]. For further increase of Cu-doping, the superconducting transition is reduced and significantly broadened. The onset $T_c$ of sample $x = 0.05$ has reached up to 20 K, exhibiting a very broad transition (~10 K). No superconductivity is observed for the Cu-doped bulks higher than 5% doping, but they undergo a metal-to-insulator transition, as depicted in the inset of Figure 5(a), comparable to the Cu-doped LiFeAs [29] and $BaFe_2As_2$ [27]. The resistivity behaviour of two pieces of sample from the same batch depicted in Figure S7 has confirmed the same superconducting behaviour, and the variation of the normal state resistivity between these two pieces could be due to the inhomogeneous distribution of the impurity phases.

In order to verify the occurrence of the Meissner effect of these samples, the temperature dependence of the magnetic moment is measured for different Cu-doped Sm1111 bulks, i.e., $x = 0, 0.01, 0.02$, and $0.05$ in both zero field cooling (ZFC) and field cooling (FC) modes in the presence of a magnetic field of 20 Oe. The normalized magnetization is depicted in Figure 6(a) in order to conduct a comparative analysis of these samples. The parent sample ($x = 0$) exhibits an onset transition of 53.2 K, which is ~1 K lower than the value obtained from the resistivity measurements. It is well known that iron-based superconductors exhibit electromagnetic granularity comparable to that of cuprates. This behaviour is also known as a weak-link behavior as well reported [42, 2]. The grain boundaries of FBS samples are observed randomly and mixed with tilt and twist grain boundaries. Recent studies have depicted that polycrystalline FBS must have a clean superconducting phase as well as well-connected grain boundaries to exhibit its high superconducting properties [43, 44, 45]. Weak grain behaviours are well observed in the magnetization studies of FBS, such as a two-step transition in the magnetic susceptibility measurements and the occurrence of two peaks during the remanent magnetization [43, 44, 45]. It is also identified as a known feature of the granular magnetic response of FBS [42, 2]. Since the resistivity behaviour is based on the percolation of the low resistive path inside the sample, the distinct influence of intergrain and intragrain behavior on the measurement of transport properties may not always be readily apparent. For example, in F-doped Sm1111, a two-step transition was observed in the magnetization measurements, but no such two-step transition was observed in the resistivity measurements, as is similar to here and consistent with earlier studies [45]. Nevertheless, it was observed that thick perovskite FBS superconductors exhibit a distinct two-step transition in both resistivity and magnetization, which can be attributed to the weak grain connection effects [46]. On this basis, the observed phenomenon of a two-step transition for $x = 0$ aligns with the previous reports [45], suggesting the well-known magnetic granular nature of iron-based superconductors [2]. It could be caused by the inhomogeneity or the existence of some non-superconducting phases between grains [42, 43, 46, 2]. The relatively low negative value of the FC magnetization suggests effective flux-pinning behavior, similar to those observed in the parent compound. The introduction of a modest quantity of Cu substitution, namely at $x = 0.01$, results in a shift from a two-step transition to a single-step transition, and this behavior persists as the Cu doping level increases. Figure 6(a) confirms the suppression of the superconducting transition and the enhancement of transition broadening with Cu doping at Fe sites, which is similar to the analysis of the resistivity behavior. Furthermore, the observed transition temperature ($T_c$) from the magnetization for different Cu-doped samples closely aligns with that observed from the transport measurements. The magnetic susceptibility measurements of two pieces of sample (Sample-1 and Sample-2) $x = 0$, 0.02, and 0.03 from the same batch are depicted in Figure S8 for Sample-1 and Sample-2, respectively, and confirm the same superconducting behavious and homogeneity of these samples.

The critical current density ($J_c$) plays an important role in the practical applications of a superconductor. The magnetic hysteresis ($M$-$H$) loop was measured for the parent sample ($x = 0$), and $x = 0.01$ bulks at 5 K for the magnetic field up to 9 T. The hysteresis loop for the parent sample is shown in the inset of Figure 6(b), and its behavior is similar to the previous reports [19]. The width of the hysteresis loop $\Delta m$ can be calculated by considering the magnetic moment in the ascending and descending magnetic fields. Generally, the Bean model is used to determine the critical current density by applying the formula: $J_c = 20\Delta m/Va(1-a/3b)$, where $a$ and $b$ are the length of the sample ($a < b$) and $V$ is the volume of the sample [47]. The rectangular-shaped samples are used for the magnetic measurements. By considering the sample dimensions, the calculated current density is depicted in Figure 6(b). The parent compound has a $J_c$ value of the order of $10^3$ A/cm$^2$, which is slightly reduced with the magnetic field up to 9 T. This behavior is similar to the previous report [18]. The 1% Cu-doped Sm1111 ($x = 0.01$) has one-order lower $J_c$ value than that of the parent compound; nevertheless, its magnetic field dependence of $J_c$ behavior is comparable to the observed for the sample $x = 0$. The low $J_c$ value suggests that the Cu-doped Sm1111 bulks have reduced pinning centers [2, 43]. It could be due to the reduced sample density and grain connections, as observed from the microstructural analysis.

To summarize our findings, the variation of the onset transition temperature ($T_c^{onset}$), the transition width ($\Delta T = T_c^{onset} - T_c^{offset}$), and the Residual Resistive Ratio ($RRR = \rho_{300K} / \rho_{60K}$) are shown in Figure 7 with the nominal Cu-doping level ($x$). These parameters are also determined from the transport measurements of two pieces of sample (Sample-1 and Sample-2) $x = 0$, 0.01, 0.02, 0.03, and 0.04 from the same batch, and are depicted in Figure S9 for Sample-1 and Sample-2. The variation of $T_c^{onset}$ parameters is also included in Figure 7(a) with respect to Ni doped SmFeAsO$_{0.88}$F$_{0.12}$ and Ru-doped SmFeAsO$_{0.85}$F$_{0.15}$. The reduction of transition temperature is almost consistent across all those transition metal substitutions. Ni substitution introduces two itinerant electrons, whereas Ru doping is isoelectronic. Based on prior observations [29], when Cu is added as a dopant, its scattering impact on electron mobility rises, resulting in an increase in resistivity (as seen in Figure 5(a)). This suggests that Cu doping acts more like an impurity center, which could be due to the localization of doping carriers. A systematic increase of the transition width ($\Delta T$) is also observed with Cu substitution, as demonstrated in Figure 7(b), which is an effect of the impurity center behavior of Cu substitution. There is a huge suppression of the $RRR$ value just for 1% Cu substitution, i.e., the $RRR$ value is reduced from 4.5 to 2 for the sample $x = 0.01$. Additionally, the substitution of Cu leads to a decrease in the $RRR$ value at a lower rate. This suggests that a small amount of Cu substitution significantly influences the presence of an impurity center, which confirms a 50% reduction in the $RRR$ value of the sample $x = 0.01$. A monotonic decrease of $RRR$ (Figure 7(c)) also reflects the evolution from the metallic state to the insulator one with Cu doping (Figure 5(a)), as consistent with previous findings on Cu-doped FeSe [30]. This phenomenon may be attributed to the Anderson localization of charge carriers and the resulting disorder caused by Cu doping [33]. Our studies support a minor electron doping effect and intact Fermi surfaces due to Cu substitution, as reported from the ARPES investigation based on Cu-doped LiFeAs [29]. These results are in contrast to other transition metals like Co and Ni doping reported for FBS [21, 20]. Therefore, rather than carrier density change, significant impurity scattering is the primary cause of $T_c$ suppression by Cu substitution. The comparison between Cu doping and Ru doping [32], on the other hand, is fascinating, as depicted in Figures 1(c)-(e). Although Cu and Ru doping are two common examples that do not include the introduction of carriers, the suppression of superconducting activity is almost comparable in both scenarios, as the behavior of $T_c$ is shown in Figure 7(a). In contrast to the case of Ru doping, where no extra carriers are induced, the Cu doping situation involves the presence of localized "carriers," suggesting a distinct physics behavior for these two substitutions. Furthermore, the determined parameters $T_c^{onset}$, $\Delta T$, and $RRR$ for two sample pieces (Sample-1 and Sample-2), as shown in Figure S9, have almost the same value with different pieces and the same behaviours with Cu doping contents, which confirm the homogeneity of the prepared Cu-doped SmFeAsO$_{0.8}$F$_{0.2}$ bulk.

Substitution or doping generally play a significant role to induce the superconductivity for high $T_c$ superconductors, i.e. iron-based superconductors [3], and the scattering potential change at the substitution sites [48]. Substitution or doping can induced disorder into crystal and produce the impurity scattering and pairbreaking effect. The presence of disorder significantly influences the phenomenon of superconductivity and the superconducting pairing gap. These effects are more interesting in the case of unconventional or multiband superconductors and depend on the symmetries and structure of superconducting order parameters. Hence, the understanding and analysis of disorder scattering phenomenon could be valuable to explore the superconducting state and mechanism [49]. Generally, the type of impurity i.e. magnetic and nonmagnetic, also plays an important role for the superconducting property. In the case a conventional superconductor, the superconducting transition temperature is not affected by the nonmagnetic impurities, following Anderson's theorem [50], whereas the magnetic impurity reduce the $T_c$ by a rate according to Abrikosov-Gor'kov theory [51]. However, the unconventional superconductor generally have a complex relation between $T_c$ reduction and the impurity scattering. The reported studies based on FBS [52, 53] suggest that the reduction of transition temperature does not strongly follow Abrikosov-Gor'kov theory [51] whether the disorder is created by magnetic [54, 49] or nonmagnetic impurity [55]. Furthermore, the intra and interband coupling scattering plays an important role to determine the precise rate of $T_c$ suppression in the multiband superconductor, where the increasing disorder reduce the transition temperature. When the scattering rate reach to the critical value, the superconductivity disappears [48] which can be a possible reason for a non-superconducting behaviour of 7% Cu doped Sm1111 ($x = 0.07$) (Figure 5). The reported studies based on Ni, Zn, and Co at Fe-sites in FBS weakly suppress $T_c$ [56, 57, 49], but the explanation based on these experimental studies is often complicated due to uncertainties in the substitutional chemistry, inhomogeneous distribution of impurities or precipitation of second phases. Impurities can also result in doping, shifting the chemical potential, all of which can mask the effect of disorder on $T_c$. Based on the previous reports [30, 29, 34], our studies suggest that Cu-doped Sm1111 has less effect on the Fermi surface due to the localized electrons and creates a disorder inside the lattice with an agreement of DFT analysis and it is consistent with the lesser contraction of the lattice parameters by copper substitution (Figure 1(c)-(e)) compared with cobalt or nickel doping [21, 20], where the itinerant electrons are induced. A reason for the reduced superconducting properties of F-doped Sm1111 could be due to strong impurity scattering instead of carrier density change as similar to the previous report for Cu doped LiFeAs [29]. Hence, the obtained results from Cu-doped Sm1111 support the existence of Cu substitutions for other iron-based superconductors [30, 29]. However, we need more advanced studies, such as ARPES based on Cu-doped Sm1111, to comprehend the Fermi energy, the scattering, and the disorder effect due to Cu substitutions, which would provide insight into the superconducting mechanisms into play.

**Conclusions**:

A series of Cu-doped SmFeAsO$_{0.8}$F$_{0.2}$ bulks was prepared using a one-step solid-state reaction method and characterized by systematic investigations through various measurements to understand the effects of Cu substitution at Fe sites on the optimally F-doped SmFeAsO. The in-plane lattice parameter '$a$' increases for Cu substitution as similar to Ni and Ru doping, whereas the out-of-plane lattice parameter '$c$' remain relatively constant for Cu doping, similar to the case of Ru substitution, but in contrast to the Ni-doped Sm1111. However, the substitution of Cu leads to an increase in the lattice volume, which is consistent with Ni and Ru substitutions where either itinerant electrons or no electrons are induced. Our observations from Raman spectroscopy and DFT calculations exhibit a strong correlation with these analyses. The Fermi surface is not significantly affected by the more localized 3d electrons from Cu dopant in Sm1111, as suggested for Cu-doped 122, 11, and 111 family. Nevertheless, the suppression rate of the onset $T_c$ is nearly identical to that reported for other transition metals (Ni, Co, and Ru) substitution. Additionally, there is an observed increase in the transition width and a reduction in the *RRR*. The observed phenomenon may be attributed to the distortion or strong impurity scattering effect inside the lattice resulting from Cu substitutions. Concisely, our study suggests that the disorder caused by the substitution of Cu metal in the superconducting layer plays a significant role in the suppression of the properties of bulk SmFeAs(O,F) superconductor. Therefore, more research is necessary to explore this aspect further.

## Acknowledgment

The research work was funded by National Science Centre (NCN), Poland, grant number "2021/42/E/ST5/00262" (SONATA-BIS 11). S.J.S. acknowledges financial support from National Science Centre (NCN), Poland through research project number: 2021/42/E/ST5/00262.

**Figure 1: (a)** X-ray diffraction patterns (XRD) of powdered SmFe$_{1-x}$Cu$_x$AsO$_{0.8}$F$_{0.2}$ ($x$ = 0, 0.01, 0.02, 0.03, 0.04, 0.05, 0.07, 0.1, and 0.2) samples at room temperature. **(b)** an enlarged view of the main peak (102) position of the parent SmFeAsO$_{0.8}$F$_{0.2}$ with respect to various Cu substitutions ($x$). The variation of **(c)** lattice parameter ($a$), **(d)** lattice parameter ($c$), and **(e)** lattice volume ($V$) with the nominal values of dopants (Cu, Ru, and Ni). The included data is taken for Ni-doped SmFeAsO$_{0.88}$F$_{0.12}$ from Ref. [21] (Y-scale: right side) and Ru-doped SmFeAsO$_{0.85}$F$_{0.15}$ from Ref. [26] (Y-scale: right side). The arrows are used to guide the eyes to depict the Y-axis scale for Ni and Ru-doped Sm1111.

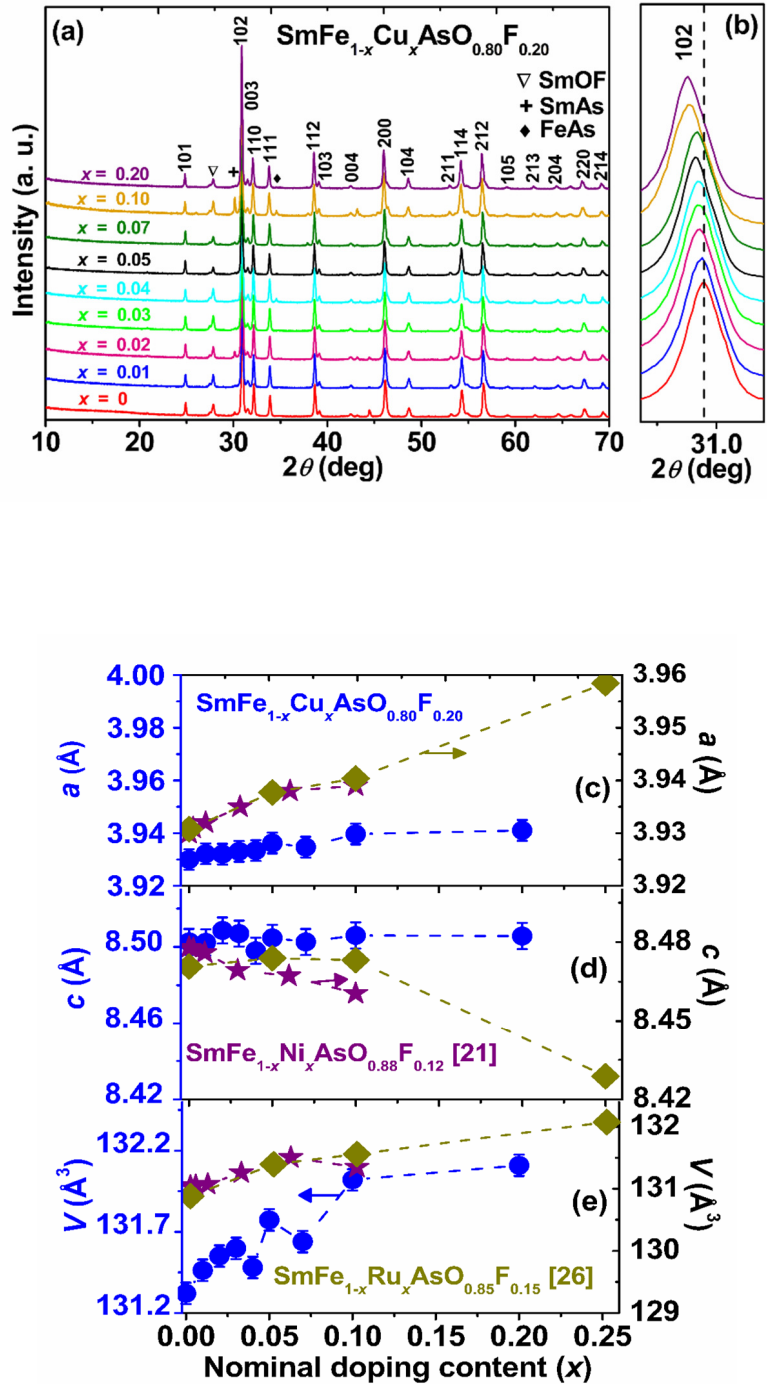

**Figure 2:** Raman spectra of Cu doped SmFeAsO$_{0.8}$F$_{0.2}$ bulks are shown. **(a)** Representative Raman spectra of the parent sample ($x = 0$) and the doped ones with copper content of 7% ($x = 0.07$) and 20% ($x = 0.20$). The depicted spectra are offset for clarity. The assignment of detected signals related to the lattice vibrations is shown for the spectrum of parent sample. The deconvolution of experimental spectra with Lorentz function is shown as green lines. Total fits of Lorentz model to experimental data are revealed by red lines. Vertical black lines are a guide for the eye. **(b)** Evolution of peak positions as a function of copper content ($X_{Cu}$). Experimental data are shown as filled circles with corresponding error bars. Dotted lines are a guide to the eye. Theoretically derived frequencies and their dependence on Cu doping are shown for Fe, As, and Sm lattice modes as solid lines of green, red, and blue color, respectively. **(c)** The calculated Fe-As distance for parent ($x = 0$), $x = 0.125$, and $x = 0.25$ samples by DFT calculation.

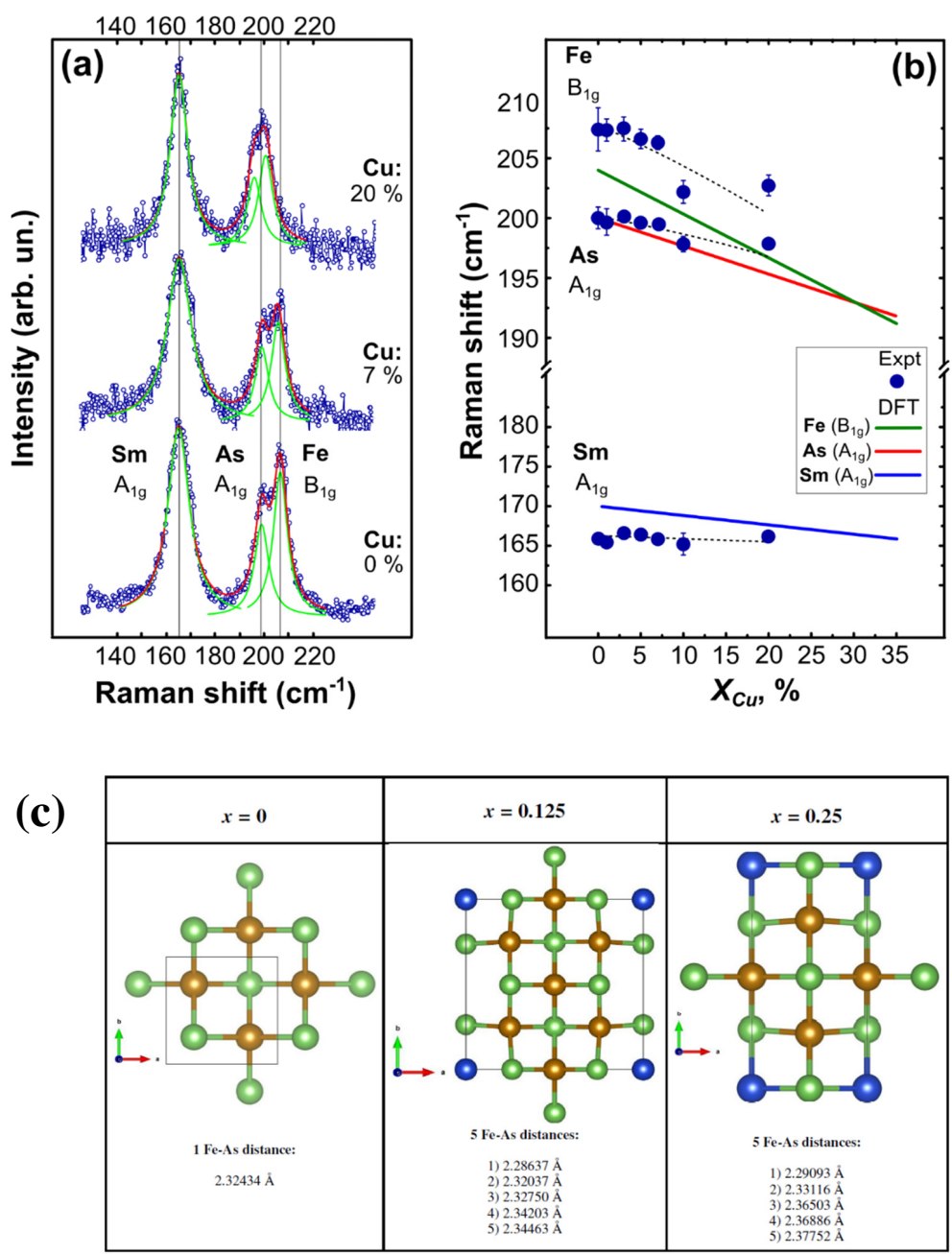

**Figure 3:** Elemental mapping for the constituent elements of SmFe$_{1-x}$Cu$_x$AsO$_{0.8}$F$_{0.2}$ polycrystalline samples: **(a)** the parent $x = 0$; **(b)** $x = 0.02$; **(c)** $x = 0.04$; **(d)** $x = 0.20$. The first and last images of each sample series are Scanning Electron Microscope (SEM) images and a combined image of all the constituent elements, respectively. The rest of the images depict the elemental mapping of the individual elements.

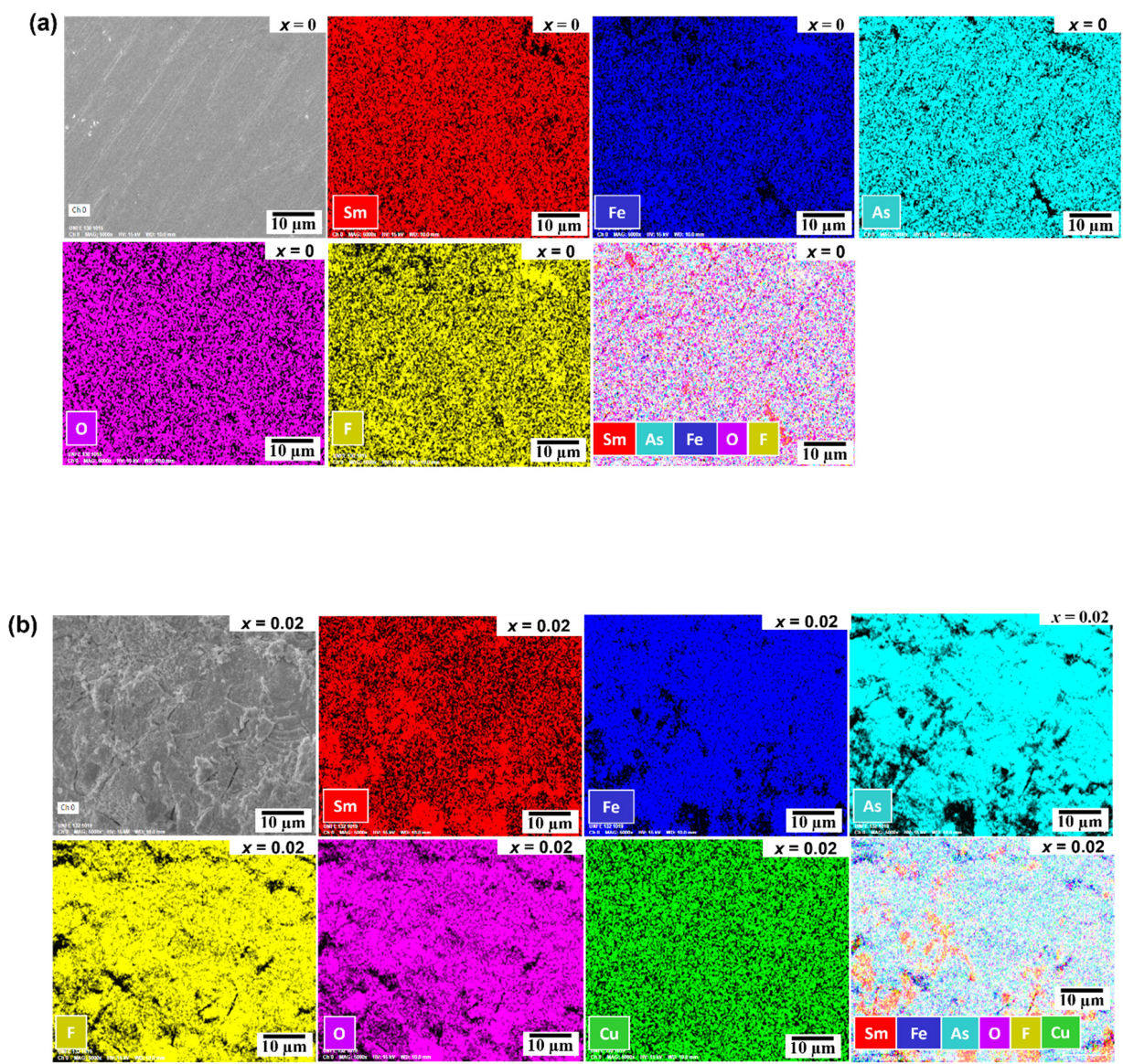

**(c)**

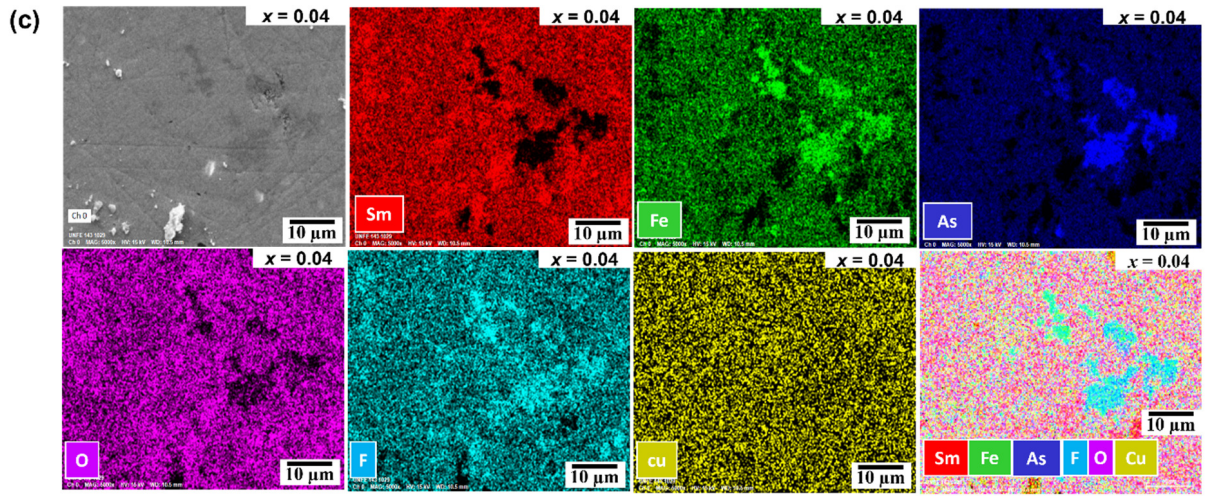

**(d)**

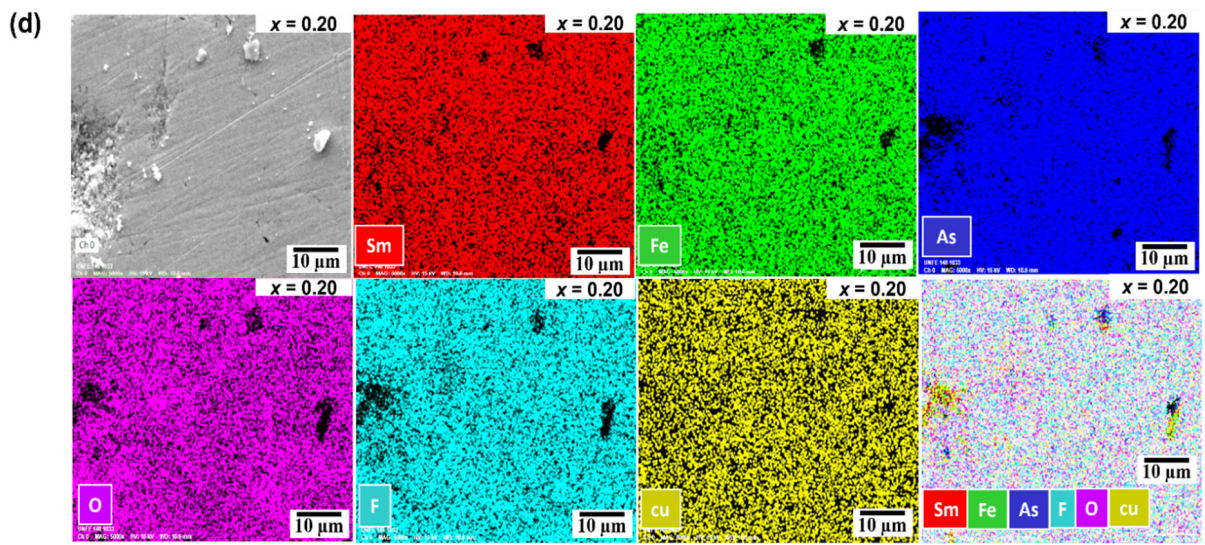

**Figure 4:** Backscattered electron image (BSE; AsB) of **(a–c)** the parent compound $x = 0$, **(d–f)** $x = 0.02$, **(g–i)** $x = 0.04$, and **(j–l)** $x = 0.2$. Bright contrast, light gray contrast, and black contrast correspond to the phases of $Sm_2O_3$, $SmFe_{1-x}Cu_xAsO_{0.8}F_{0.2}$, and pores (which occasionally can be SmAs/FeAs), respectively.

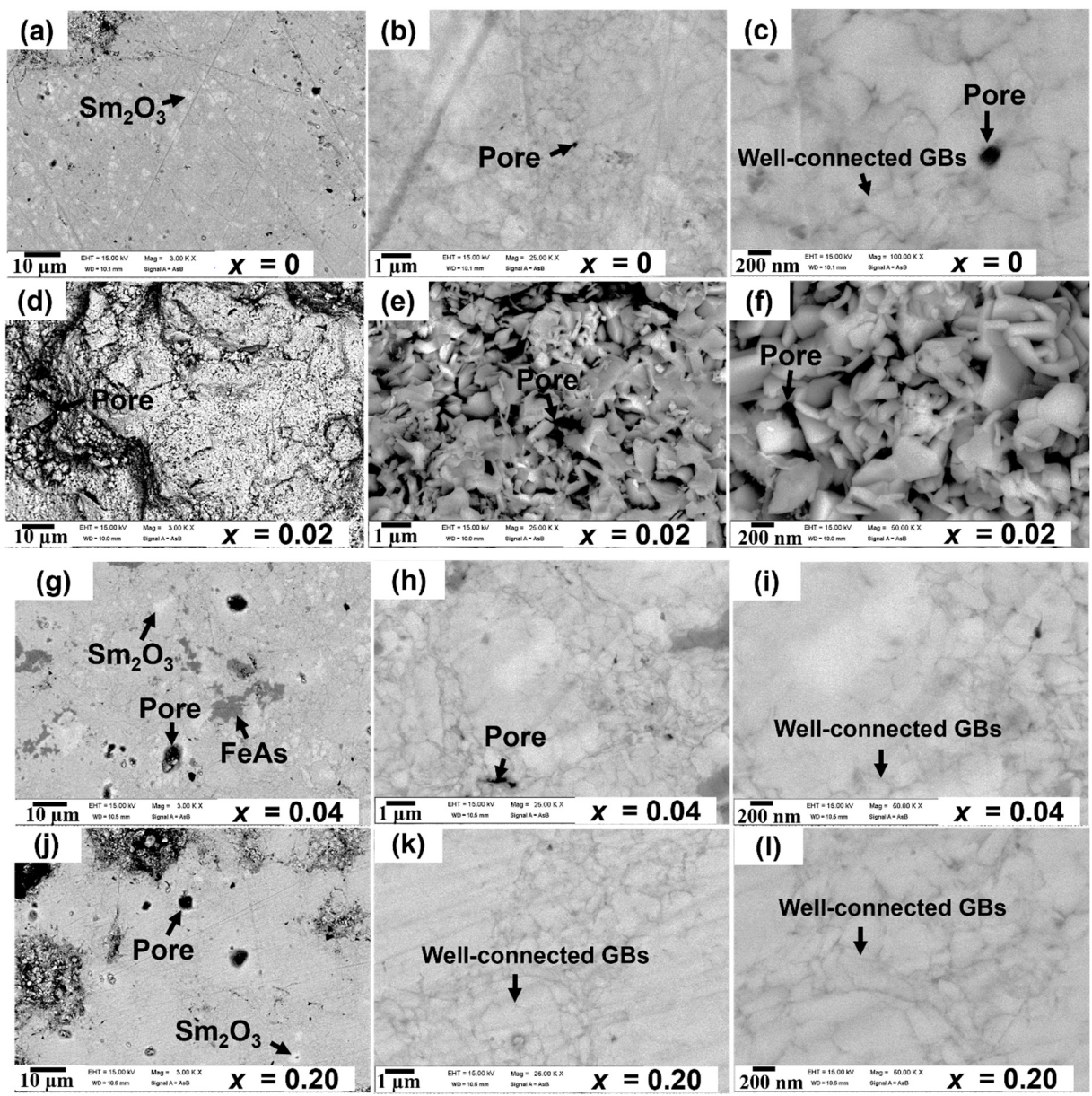

**Figure 5:** **(a)** The variation of resistivity ($\rho$) with the temperature up to the room temperature for SmFe$_{1-x}$Cu$_x$AsO$_{0.8}$F$_{0.2}$. The inset figure shows the temperature dependence of the resistivity for the sample $x = 0.07$, 0.1, and 0.2. **(b)** Low-temperature variation of the resistivity up to 60 K for various Cu-doped SmFeAsO$_{0.8}$F$_{0.2}$. The inset figure illustrates a method for defining the onset ($T_c^{onset}$) and offset ($T^{offset}$) superconducting transitions.

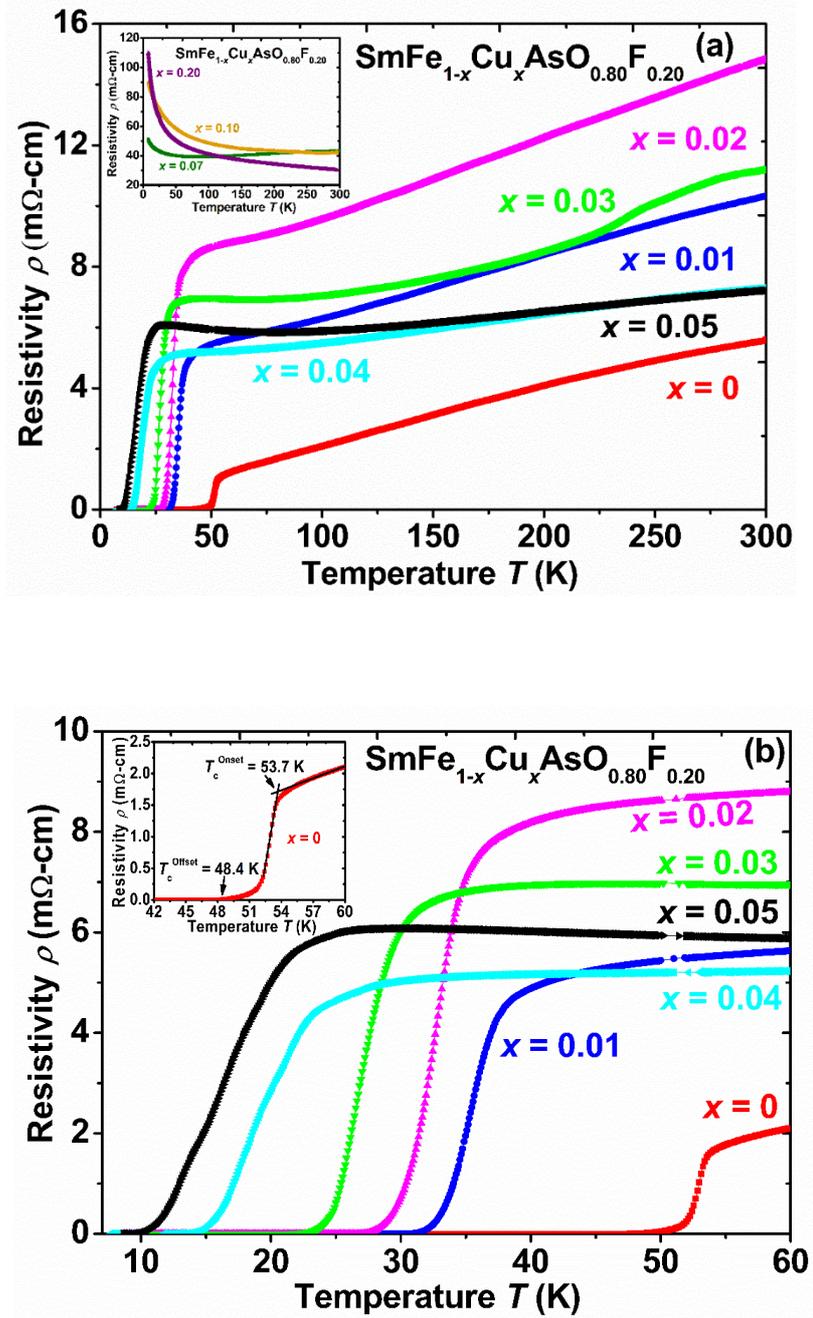

**Figure 6: (a)** The temperature dependence of the normalized magnetic moment for SmFe$_{1-x}$Cu$_x$AsO$_{0.8}$F$_{0.2}$ ($x$ = 0, 0.01, 0.02, 0.03, 0.05) in ZFC and FC mode under a magnetic field of 20 Oe. **(b)** The variation of the critical current density ($J_c$) with the applied magnetic field for $x$ = 0 and 0.01 at a temperature of 5 K and a magnetic field of up to 9 T. The inset of Figure 6(b) illustrates the magnetic hysteresis loop ($M$-$H$) for the parent ($x$ = 0) sample at a temperature of 5 K, in the presence of the magnetic field up to 9 T.

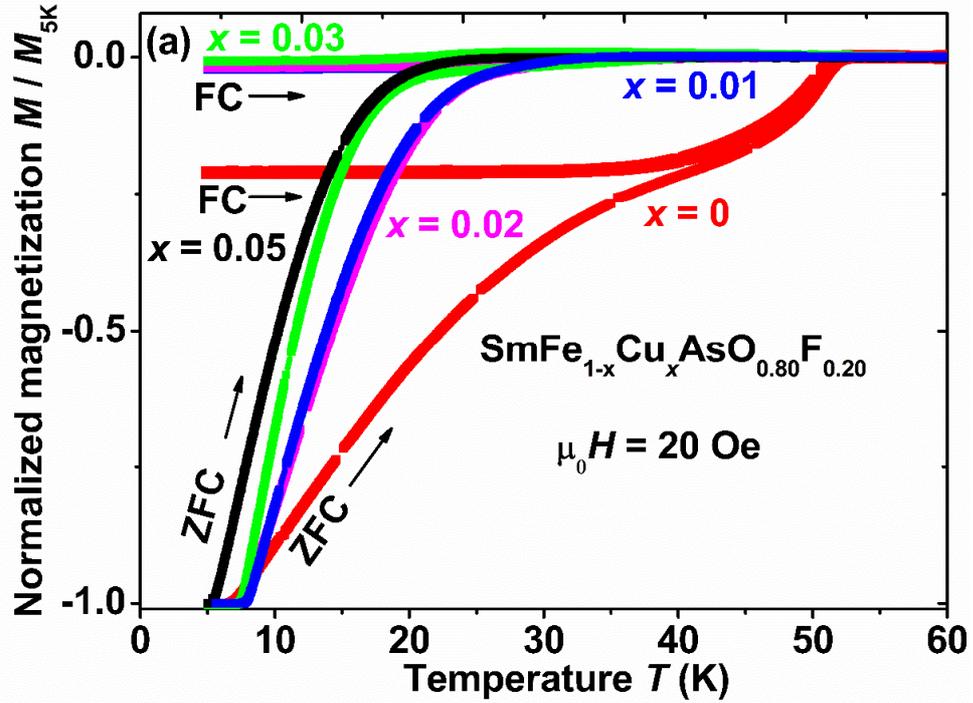

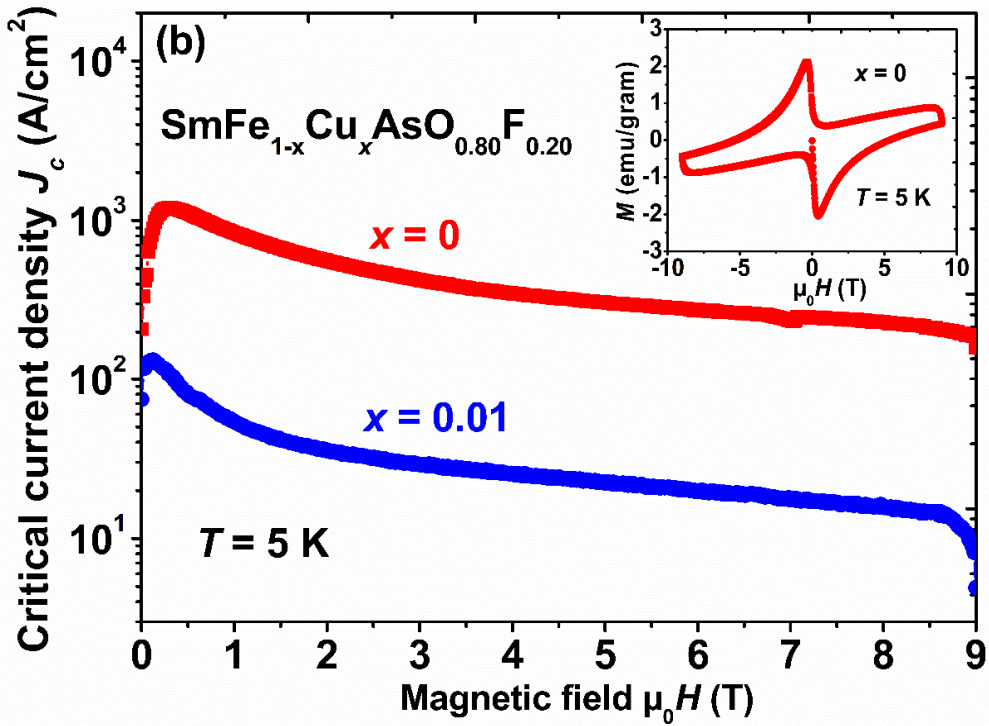

**Figure 7:** The variation of **(a)** the onset transition temperature ($T_c^{onset}$), **(b)** the transition width ($\Delta T = T_c^{onset}$ - $T_c^{offset}$), and **(c)** $RRR$ (= $\rho_{300\,K}$ / $\rho_{60\,K}$) of SmFe$_{1-x}$Cu$_x$AsO$_{0.8}$F$_{0.2}$ with the nominal doping contents ($x$). In Figure (a), the reported $T_c^{onset}$ data is also included for Ni-doped SmFeAsO$_{0.88}$F$_{0.12}$ from Ref. [21] and Ru-doped SmFeAsO$_{0.85}$F$_{0.15}$ from Ref. [26]. Based on the measured two pieces of each sample from the same batch, the variation of $T_c^{onset}$, $\Delta T$, and $RRR$ with Cu-doping content ($x$) is shown in Figure S9. Following that, the calculated error bars for $T_c^{onset}$, $\Delta T$, and $RRR$ are around 3%, 6%, and 4%, which are included in Figures 7(a), 7(b), and 7(c), respectively, for our Cu-doped SmFeAsO$_{0.8}$F$_{0.2}$.

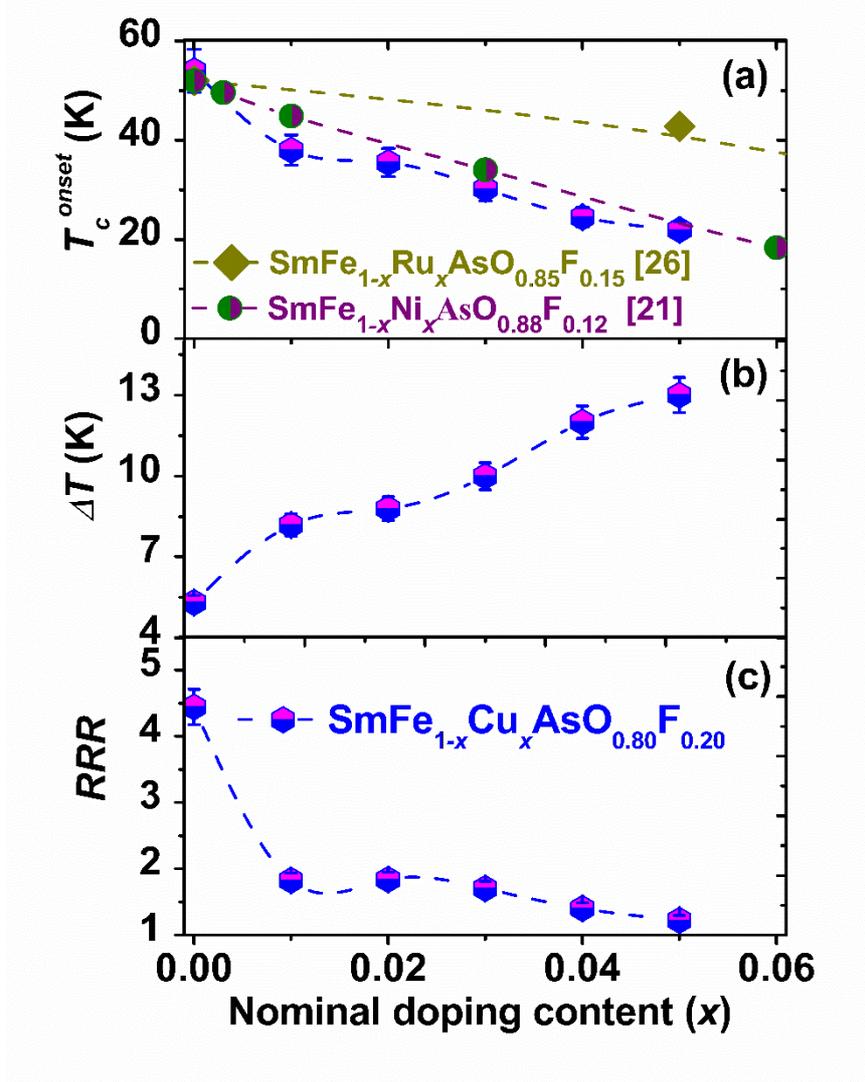

# Copper doping effects on the superconducting properties of Sm-based oxypnictides

Mohammad Azam[1], Manasa Manasa[1], Tatiana Zajarniuk[2], Taras Palasyuk[1], Ryszard Diduszko[3], Tomasz Cetner[1], Andrzej Morawski[1], Cezariusz Jastrzebski[4], Michał Wierzbicki[4], Andrzej Wiśniewski[2], Shiv J. Singh[1*]

[1]*Institute of High Pressure Physics (IHPP), Polish Academy of Sciences, Sokołowska 29/37, 01-142 Warsaw, Poland*

[2]*Institute of Physics, Polish Academy of Sciences, aleja Lotników 32/46, 02-668 Warsaw, Poland*

[3]*Łukasiewicz Research Network Institute of Microelectronics and Photonics, Aleja Lotników 32/46, 02-668 Warsaw, Poland*

[4]*Faculty of Physics, Warsaw University of Technology, Koszykowa 75, 00-662 Warsaw, Poland*

*\*Correspondence address: sjs@unipress.waw.pl*

*https://orcid.org/0000-0001-5769-1787*

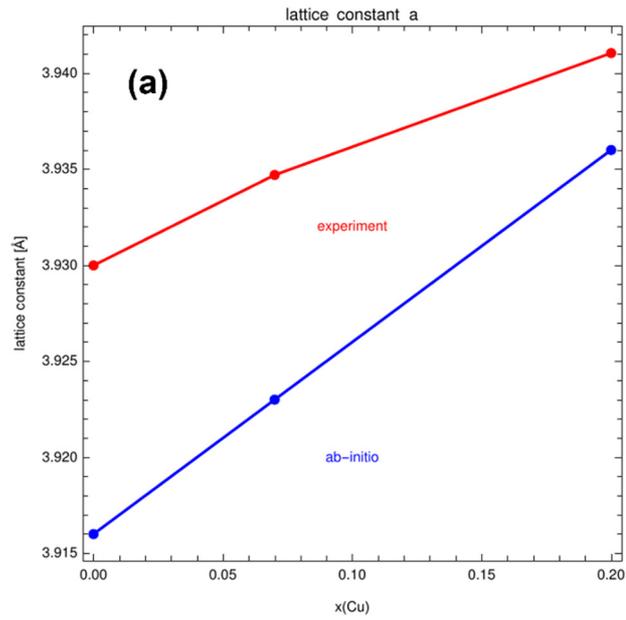

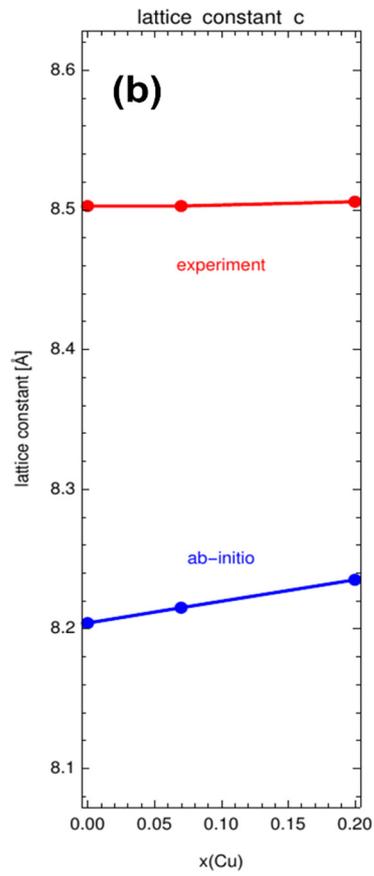

**Figure S1:** The variation of experimental lattice parameters *a* and *c* obtained from the structural analysis (experiment) and the calculated parameters from DFT calculations (ab-initio) is shown with the nominal Cu content (*x*) for SmFe$_{1-x}$Cu$_x$AsO$_{0.8}$F$_{0.2}$ bulks.

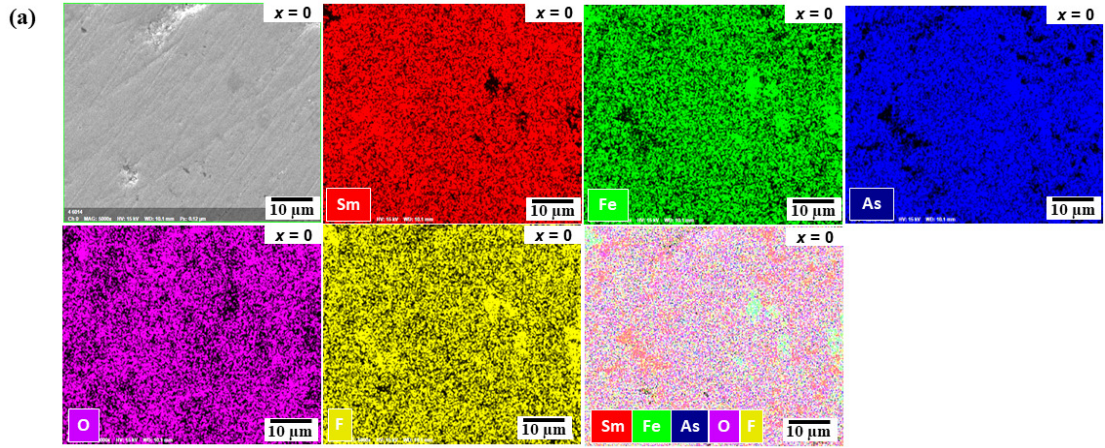

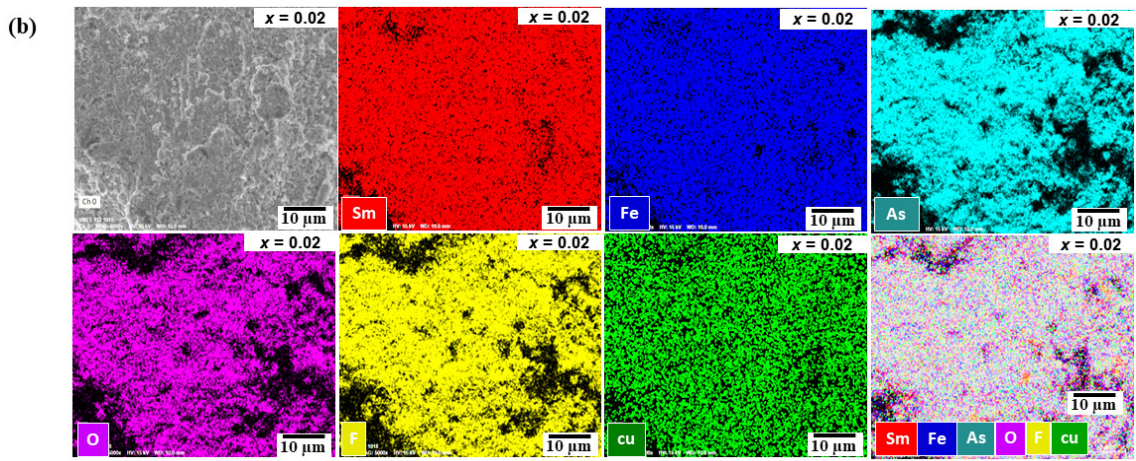

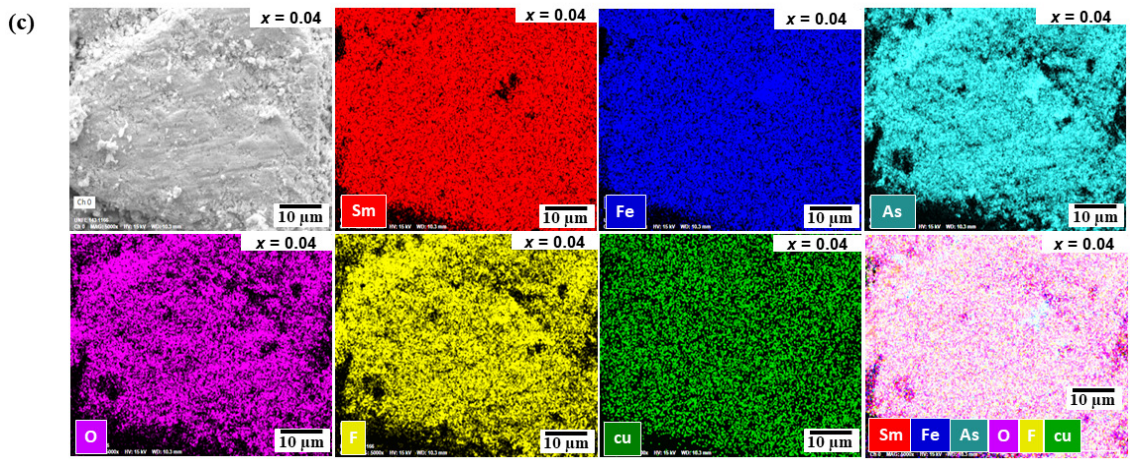

**(d)**

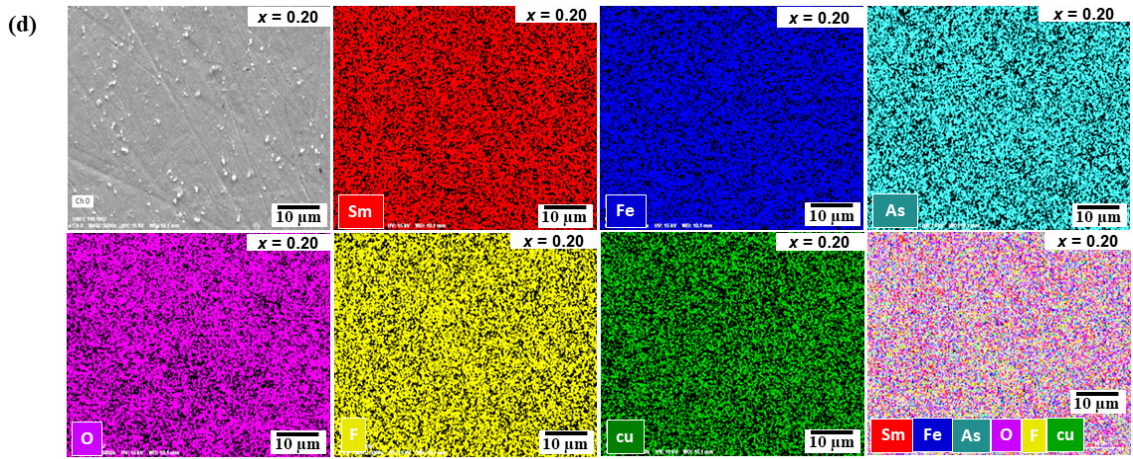

**Figure S2:** The elemental mapping for the constituent elements of $SmFe_{1-x}Cu_xAsO_{0.8}F_{0.2}$ polycrystalline samples: (**a**) $x = 0$, (**b**) $x = 0.02$, (**c**) $x = 0.04$, and (**d**) $x = 0.20$. The depicted measurements, here, are performed for Sample-2, whereas the respective measurements for Sample-1 are shown in Figure 3. One can note that Sample-1 and Sample-2 are two sample pieces from the same batch sample $x = 0$, 0.02, 0.04 and 0.20. The first and last images of each sample series are SEM images and a combined image of all the constituent elements, respectively. The rest of the images depict the elemental mapping of the individual elements.

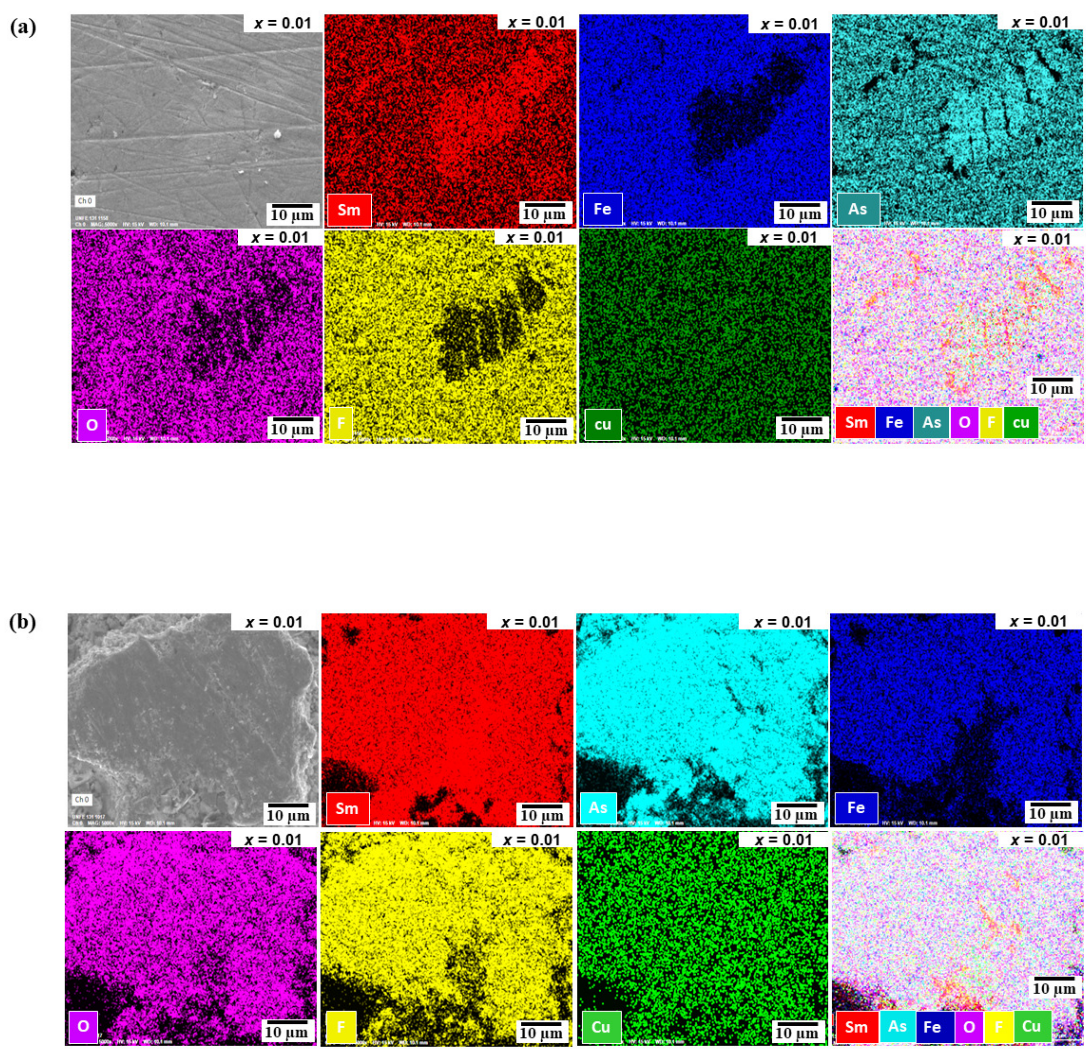

**Figure S3:** The elemental mapping for the constituent elements of $SmFe_{0.99}Cu_{0.01}AsO_{0.8}F_{0.2}$ ($x = 0.01$) polycrystalline samples for **(a)** Sample-1 and **(b)** Sample-2. These samples are two pieces from the same batch of each sample. The first and last images of each sample series are SEM images and a combined image of all the constituent elements, respectively. The rest of the images depict the elemental mapping of the individual elements.

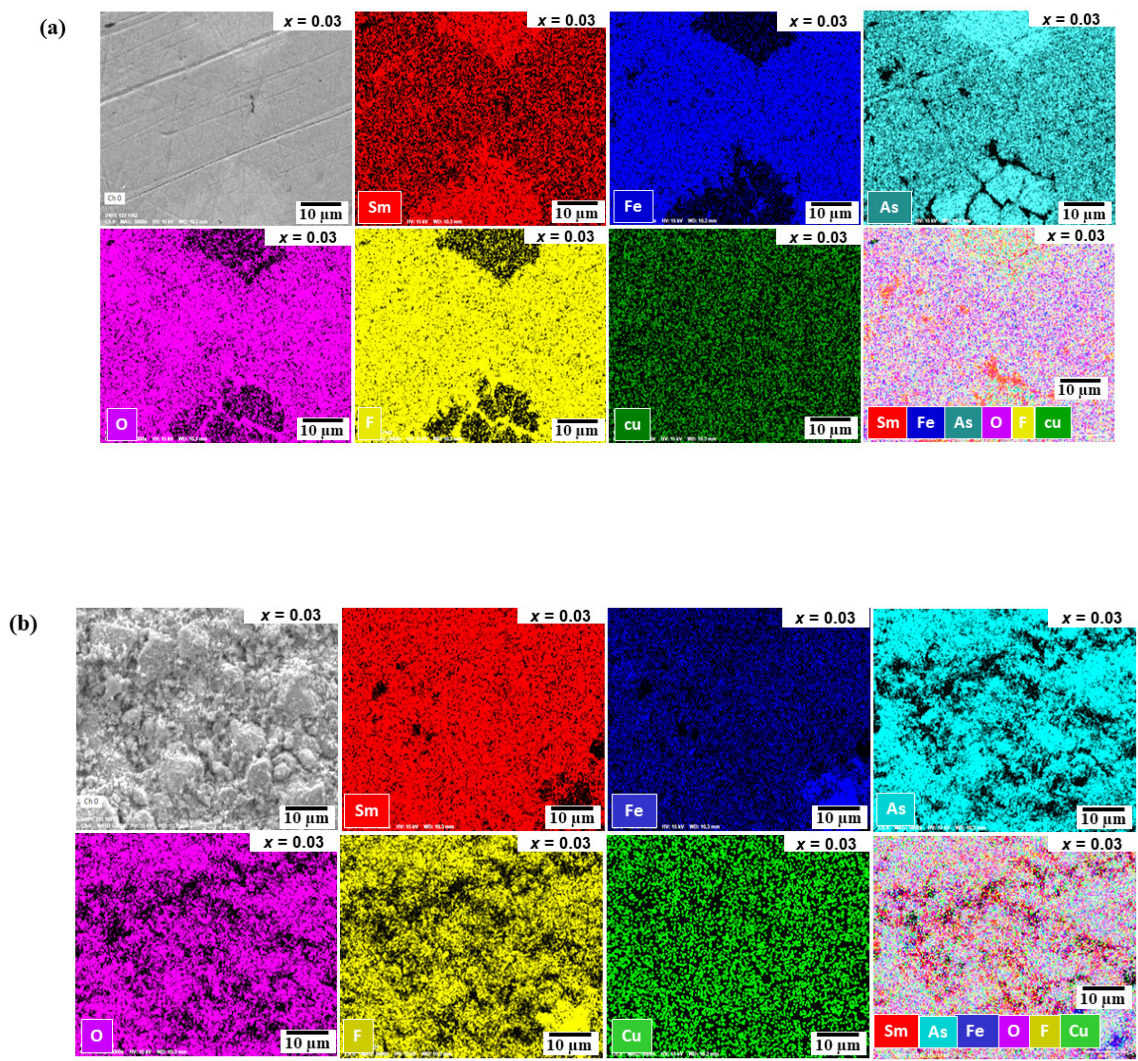

**Figure S4:** The elemental mapping for the constituent elements of $SmFe_{0.97}Cu_{0.03}AsO_{0.8}F_{0.2}$ ($x = 0.03$) polycrystalline samples for **(a)** Sample-1 and **(b)** Sample-2. These samples are two pieces from the same batch of each sample. The first and last images of each sample series are SEM images and a combined image of all the constituent elements, respectively. The rest of the images depict the elemental mapping of the individual elements.

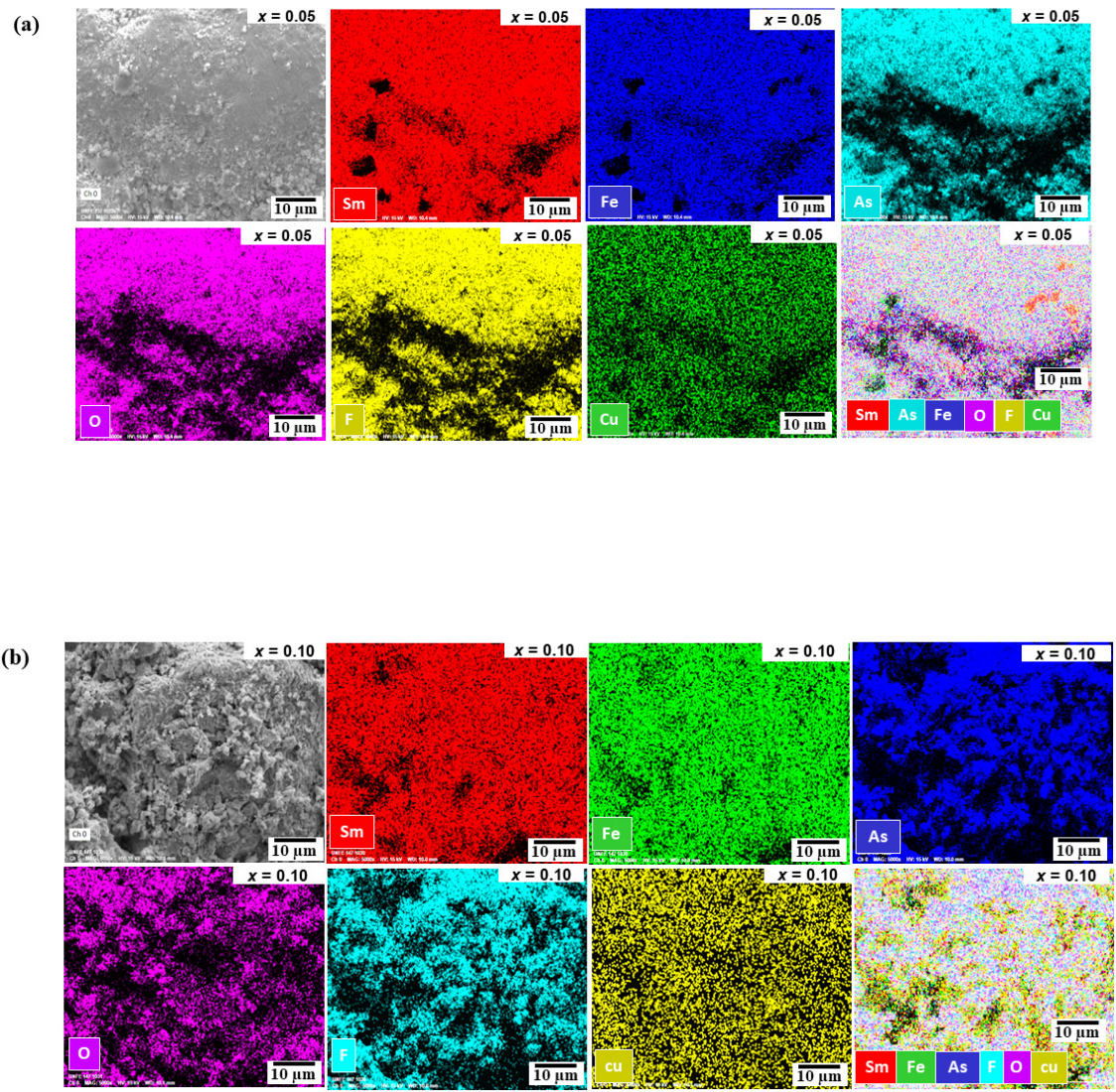

**Figure S5:** The elemental mapping for the constituent elements of **(a)** SmFe$_{0.95}$Cu$_{0.05}$AsO$_{0.8}$F$_{0.2}$ **(b)** SmFe$_{0.90}$Cu$_{0.10}$AsO$_{0.8}$F$_{0.2}$ polycrystalline samples. The first and last images of each sample series are SEM images and a combined image of all the constituent elements, respectively. The rest of the images depict the elemental mapping of the individual elements.

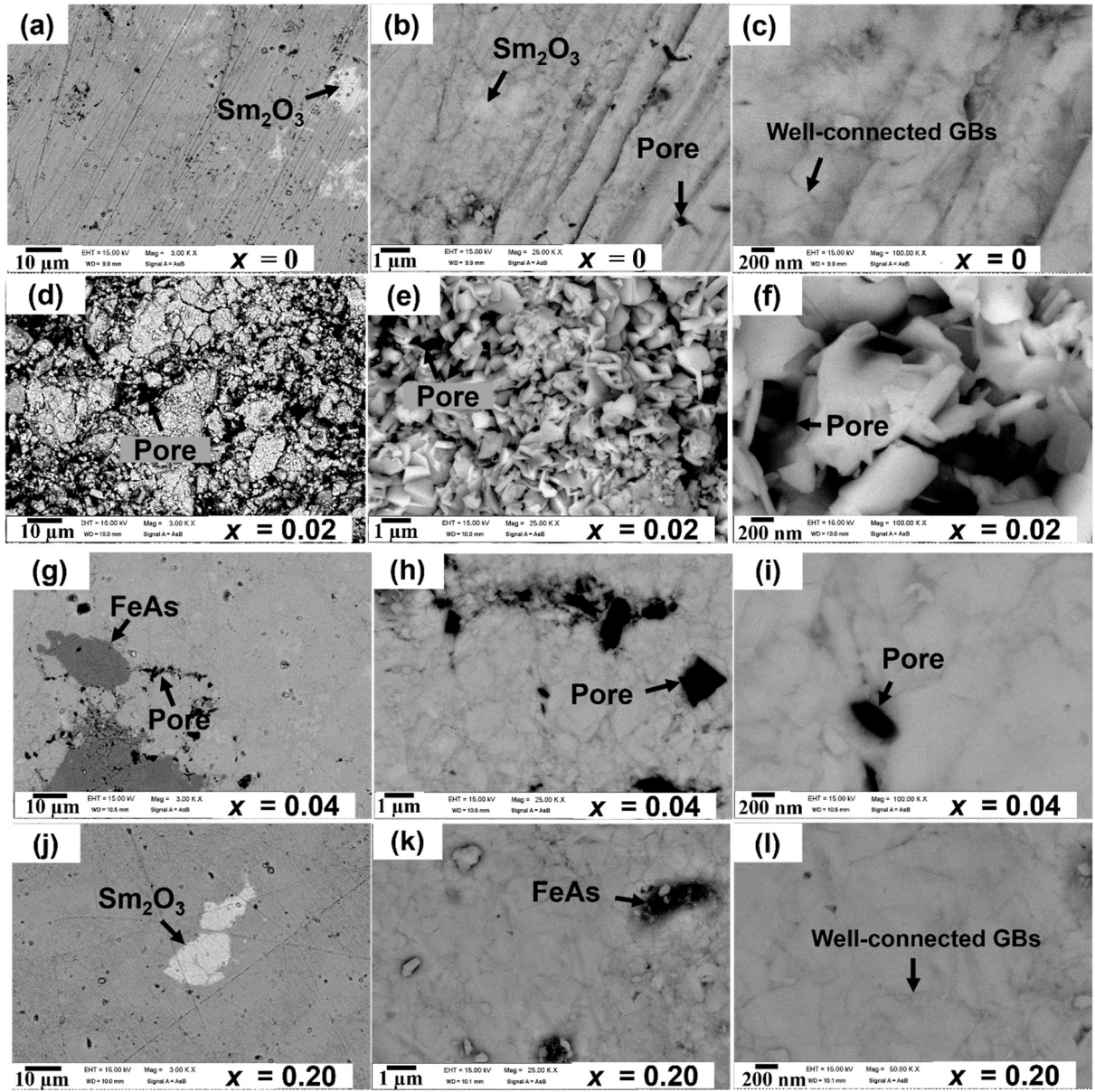

**Figure S6:** Backscattered electron image (BSE; AsB) of SmFe$_{1-x}$Cu$_x$AsO$_{0.8}$F$_{0.2}$ bulks for **(a–c)** the parent sample $x = 0$, **(d–f)** $x = 0.02$, **(g–i)** $x = 0.04$, and **(j–l)** $x = 0.20$ performed on "Sample-2". Bright contrast, light gray, and black contrast correspond to the phases of Sm$_2$O$_3$ (SmOF), SmFe$_{1-x}$Cu$_x$AsO$_{0.8}$F$_{0.2}$, and pores (FeAs/SmAs), respectively. The images depicted in Figure 4 are based on Sample-1. Sample-1 and Sample-2 are two pieces from the same batch of each Cu-doped Sm1111.

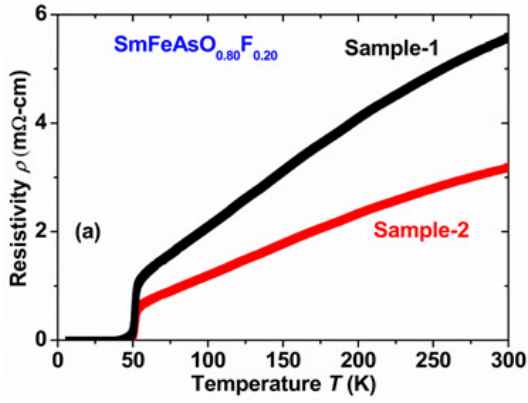

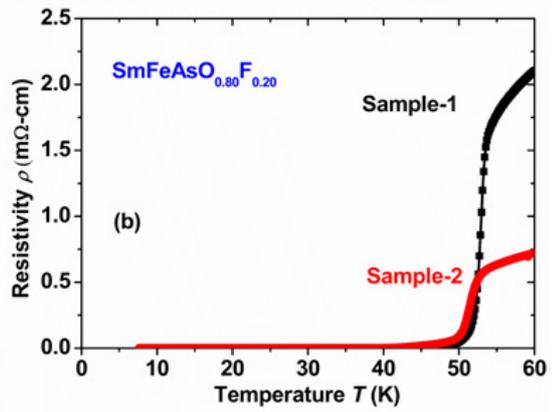

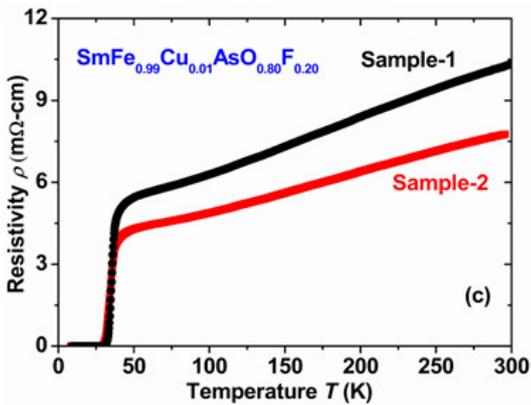

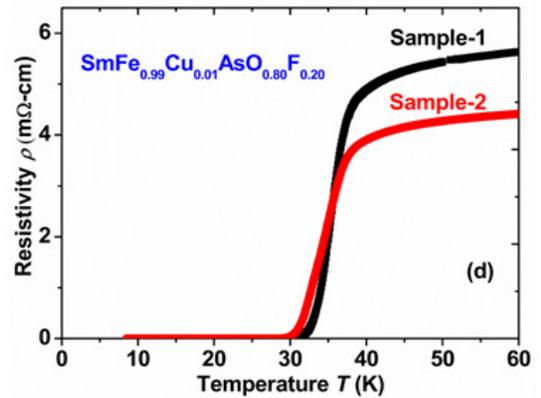

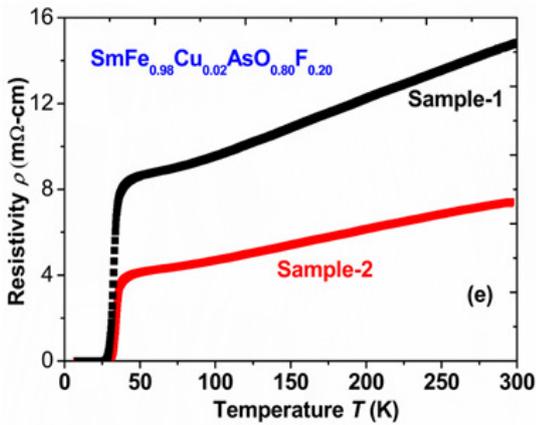

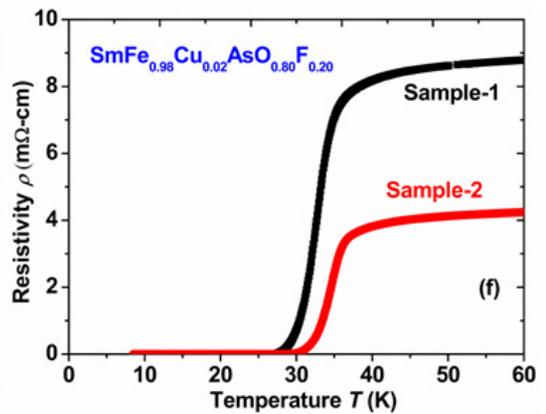

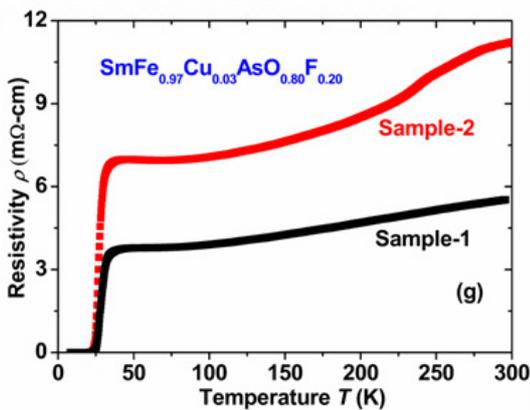

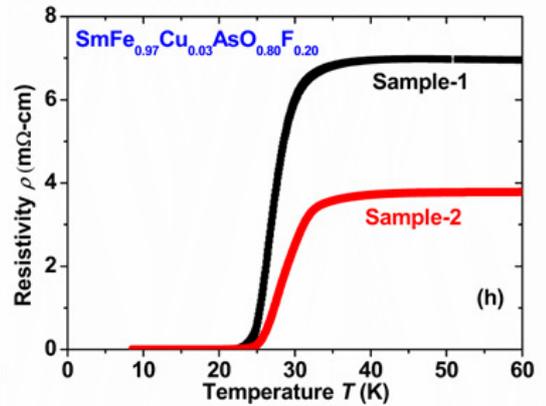

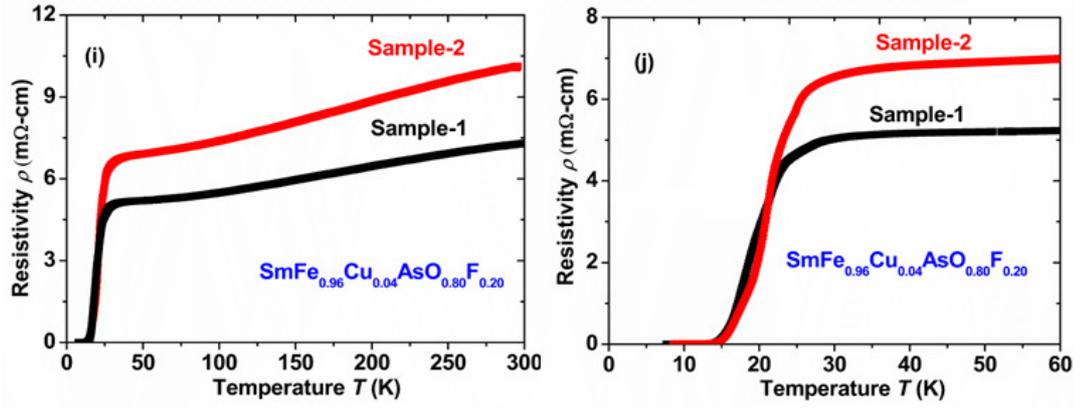

**Figure S7:** The temperature dependence of the resistivity ($\rho$) up to the room temperature for **(a)** $x = 0$, **(c)** $x = 0.01$, **(e)** $x = 0.02$, **(g)** $x = 0.03$, and **(i)** $x = 0.04$ for sample-1 and sample-2. Low-temperature variation of the resistivity up to 60 K for various $SmFe_{1-x}Cu_xAsO_{0.8}F_{0.2}$ : **(b)** $x = 0$, **(d)** $x = 0.01$, **(f)** $x = 0.02$, **(h)** $x = 0.03$, and **(j)** $x = 0.04$ for sample-1 and sample-2. One can note that Sample-1 and Sample-2 are two pieces from the same batch of each Cu-doped Sm1111 bulks.

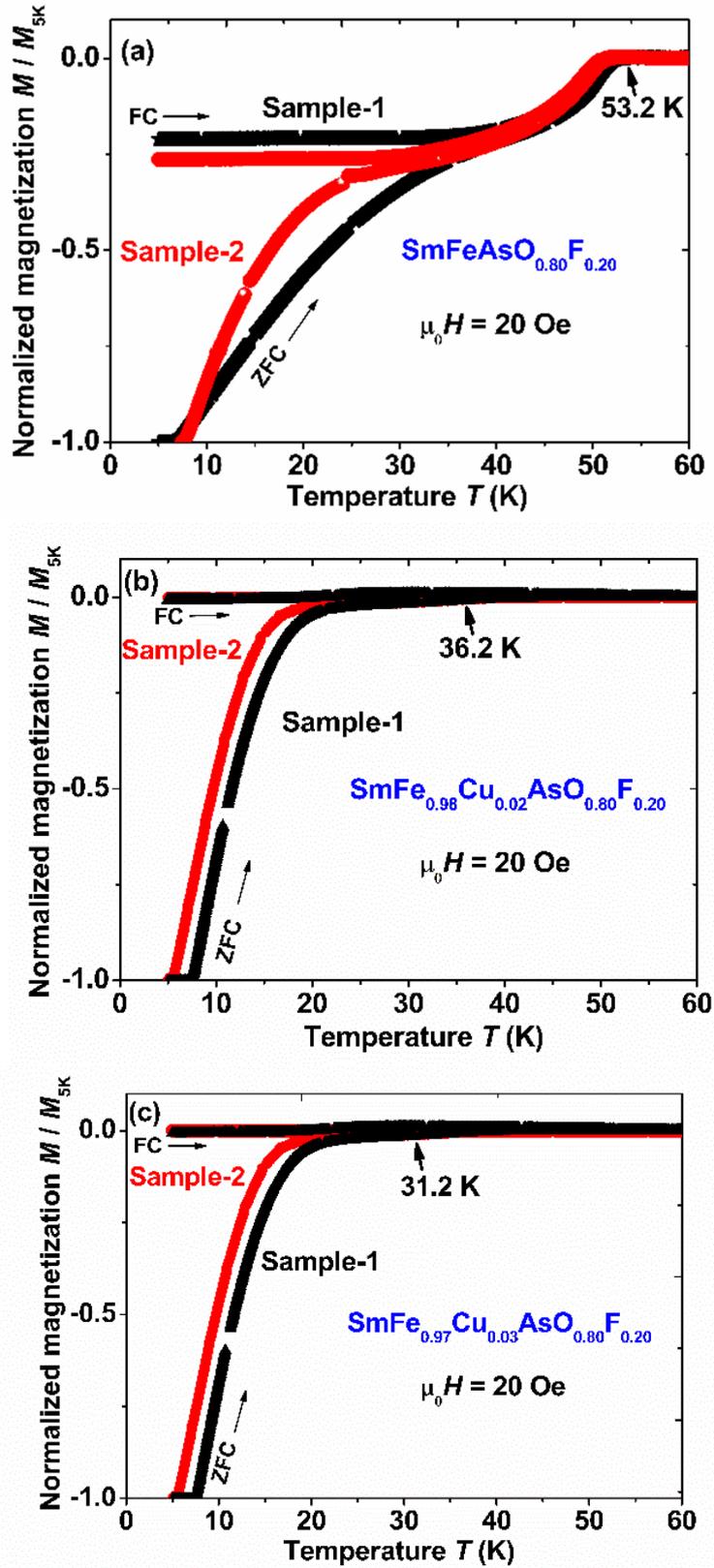

**Fig. S8.** The variation of the normalized magnetic moment ($M / M_{5K}$) with the temperature in the presence of a magnetic field of 20 Oe for SmFe$_{1-x}$Cu$_x$AsO$_{0.8}$F$_{0.2}$ **(a)** $x = 0$, **(b)** $x = 0.02$, and **(c)** $x = 0.03$, respectively, for sample-1 and sample-2, which are two pieces from the same batch of each Cu-doped Sm1111 bulk.

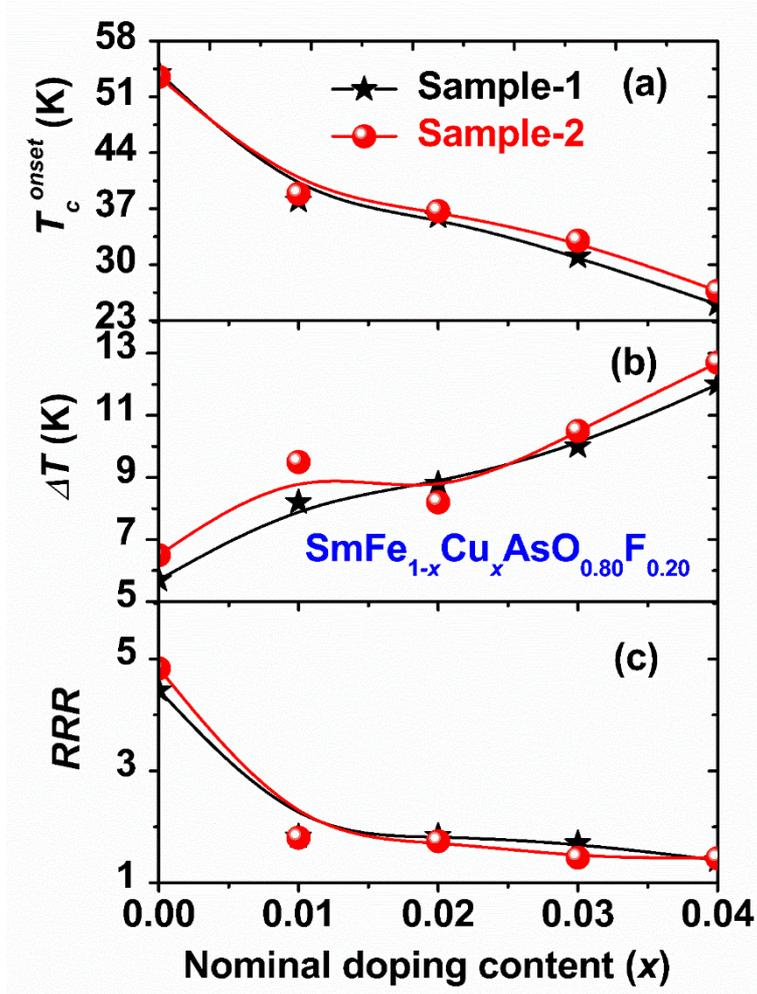

**Figure S9 : (a)** The onset transition temperature $T_c^{onset}$, **(b)** the transition width ($\Delta T$), and **(c)** $RRR$ (= $\rho_{300\,K}$ / $\rho_{60\,K}$) variation with Cu-doping content ($x$) for SmFe$_{1-x}$Cu$_x$AsO$_{0.8}$F$_{0.2}$ bulks for sample-1 and sample-2. These samples (Sample-1 and Sample-2) are two pieces from the same batch of each Cu-doped Sm1111 bulk.